\documentclass{aa}  

\usepackage{graphicx}
\usepackage{txfonts}

\usepackage{natbib,twoopt}

\begin{document}

   \title{Surveying the Whirlpool at Arcseconds with NOEMA (SWAN)}

   \subtitle{IV. Extent of active galactic nucleus feedback on the interstellar medium}

    \author{Mallory D. Thorp\inst{\ref{AlfA}}
        \and Antonio Usero\inst{\ref{OAN}}
        \and Frank Bigiel\inst{\ref{AlfA}}
        \and Ina Galić\inst{\ref{AlfA}}
        \and Smita Mathur \inst{\ref{OSU},\ref{OSU_C}}
        \and Sophia K. Stuber\inst{\ref{NAOJ},\ref{MPIfR} } \thanks{International Research Fellow of Japan Society for the Promotion of Science (Postdoctoral Fellowships for Research in Japan)}
        \and Jennifer A. Rodriguez \inst{\ref{OSU}, \ref{OSU_C}}
        \and Dario Colombo \inst{\ref{AlfA}}
        \and Bret Lehmer \inst{\ref{UArk},\ref{Ark}}
        \and Eva Schinnerer \inst{\ref{MPIA}}
        \and Amirnezam Amiri \inst{\ref{UArk}}
        \and Ashley Barnes \inst{\ref{ESO}}
        \and Zein Bazzi \inst{\ref{AlfA}}
        \and Guillermo A. Blanc \inst{\ref{OCIS},\ref{UC}}
        \and Cosima Eibensteiner \inst{\ref{NRAO}}\thanks{Jansky Fellow of the National Radio Astronomy Observatory}
        \and Simon Glover \inst{\ref{ITA}}
        \and Laura A. Lopez \inst{\ref{OSU},\ref{OSU_C}}
        \and Erik B. Monson \inst{\ref{MTSU}}
        \and Lukas Neumann \inst{\ref{ESO}}
        \and Jerome Pety \inst{\ref{IRAM},\ref{LUX}}
        \and Miguel Querejeta \inst{\ref{OAN}}
        \and Thomas G. Williams\inst{\ref{Manc}}
}
    \institute{
    Argelander-Institut für Astronomie, Universität Bonn, Auf dem Hügel 71, 53121 Bonn, Germany\label{AlfA}
    \and 
    Observatorio Astron\'omico Nacional (IGN), C/ Alfonso XII 3, E-28014 Madrid, Spain\label{OAN}
    \and 
    Department of Astronomy, The Ohio State University, 140 West 18th Ave, Columbus, OH 43210, USA\label{OSU}
    \and
    Center for Cosmology and Astro-Particle Physics, Ohio State University, Columbus, OH, USA\label{OSU_C}
    \and 
    National Astronomical Observatory of Japan, 2-21-1 Osawa, Mitaka, Tokyo 181-8588, Japan\label{NAOJ}
    \and 
    Max Planck Institute for Radio Astronomy, Auf dem Hügel 69, 53121 Bonn, Germany\label{MPIfR}
    \and 
    Max-Planck-Institut für Astronomie, Königstuhl 17, 69117 Heidelberg Germany\label{MPIA}
    \and
    Department of Physics, University of Arkansas, 226 Physics Building, 825 West Dickson Street, Fayetteville, AR 72701, USA\label{UArk}
    \and
    Arkansas Center for Space and Planetary Sciences, University of Arkansas, 332 N. Arkansas Ave, Fayetteville, AR 72701, USA\label{Ark}
    \and
    European Southern Observatory (ESO), Karl-Schwarzschild-Straße 2, 85748 Garching, Germany\label{ESO}
    \and
    Observatories of the Carnegie Institution for Science, 813 Santa Barbara Street, Pasadena, CA 91101, USA\label{OCIS}
    \and
    Departamento de Astronom\'{i}a, Universidad de Chile, Camino del Observatorio 1515, Las Condes, Santiago, Chile\label{UC}
    \and
    National Radio Astronomy Observatory, 520 Edgemont Rd, Charlottesville, VA 22903, USA\label{NRAO}
    \and 
    Institute for Theoretical Astrophysics, Albert-Ueberle-Strasse 2, 69120 Heidelberg\label{ITA}
    \and
    Department of Physics and Astronomy, Middle Tennessee State University, 1301 E. Main St Box 71, Murfreesboro, TN 37312\label{MTSU}
    \and
    IRAM, 300 rue de la Piscine, F-38406 Saint Martin d’Hères, France\label{IRAM}
    \and
    LUX, Observatoire de Paris, PSL Research University, CNRS, Sorbonne Universités, 75014 Paris, France\label{LUX}
    \and
    UK ALMA Regional Centre Node, Jodrell Bank Centre for Astrophysics, Department of Physics and Astronomy, The University of Manchester, Oxford Road, Manchester M13 9PL, UK\label{Manc}
}

   \date{Received December 19 2025; accepted March 25 2026}
 
\abstract 
   {Active Galactic Nuclei (AGN) are intertwined with galaxy evolution, injecting energy into the interstellar medium (ISM) and possibly regulating star formation as a galaxy evolves. However, the phenomena through which we observe AGN are multiphase and multiscale, which can lead to conflicting results for how significantly and to what extent AGN influence the ISM.} 
   {M51 is a perfect case study of the boundary between where AGN feedback and star formation feedback dominate the ISM, hosting a low-luminosity type II Seyfert nucleus with a well-defined molecular and ionized outflow. We endeavor to characterize the spatial extent and dominant modes of AGN feedback in M51 utilizing multiple phases of the ISM.}
   {Using integral field spectroscopy observations from VENGA of the central 3 kpc, we identified regions dominated by AGN ionization using an emission line ratio (ELR) function. We then combined this information with new observations of the dense molecular ISM in M51 from SWAN, including cloud-scale mapping of HCN(1--0), HNC(1--0), HCO$^+$(1--0), and N$_2$H$^+$(1--0). Both datasets allowed us to achieve $\sim$180\,pc resolution, allowing for a clear demarcation of where AGN feedback dominates the ISM. We then tested how the ELR compares to other tracers of AGN activity, using both millimeter emission line ratios as well as X-ray observations from Chandra to assess the dominant mode of feedback.}
   {If we assume that N$_2$H$^+$(1--0) is the best tracer of dense, cold gas in SWAN, then AGN-dominated regions defined by the ELR all have greater emission in (1--0) transitions in HCN, HNC, and HCO$^+$ than would be expected if they traced dense gas alone, implying excitation of these lines from AGN feedback. The ELR is better at selecting these regions compared to molecular tracers of AGN activity, such as HCN(1--0)/HCO$^+$(1--0), which are heightened for a greater extent in M51. Some of the highest ELR values are also associated with fast shocks evident in the optical, which are concurrent with large  HNCO(4--3)/CO(1--0) values that point to slow shocks near the nucleus. The presence of shocks and heightened N$_2$H$^+$(1--0) near the nucleus indicate a potential dense molecular outflow, meaning heightened dense tracer emission could be partly due to larger abundance rather than excitation alone in this region.}
   {All tracers of AGN activity point to a ``two-stage'' feedback scenario, whereby mechanical feedback from the jet-ISM interaction spurs soft X-ray emission that excites molecules such as HCN. Dense gas entrenched in a molecular outflow may also lead to a greater chemical abundance of multiple tracers measured with SWAN, but to a lesser extent than excitation from AGN feedback.}


   \maketitle

\section{Introduction}
\label{sec:intro}
Active galactic nuclei (AGNs) play a key role in the regulation and evolution of star formation in massive galaxies. Contemporary theoretical works point to AGNs as an essential modulator for galaxy growth and the cessation of star formation, requiring some form of AGN feedback to reproduce observed galaxy populations \citep{Somerville2015SouthSciences,Leslie2016QuenchingSequence,Husemann2018RealityFeedback}. This means that AGN feedback must have some effect on the properties and distribution of the interstellar medium (ISM) to regulate star formation. However, the extent of the impact AGNs have on the ISM is still a point of contention based on observational signatures, resulting from the variety of multiwavelength phenomena that arise from nuclear activity \citep{Fabian2012ObservationalFeedback,Padovani2017ActiveName,Hickox2018ObscuredNuclei}. Not all AGNs exhibit every component across the electromagnetic spectrum, with the strength of AGNs, recency of activity, and the orientation of a jet and/or outflow all influencing detectability. Characterizing the extent and magnitude of different types of AGN activity on the ISM is thus crucial to disentangling the variety of processes that regulate star formation across a galaxy's lifetime.

Active galactic nuclei inject both kinetic and radiative energy into the ISM, exciting gas and creating shocks through multi-phase outflows and relativistic jets, which leave a variety of signatures over a period of time. Kinetic feedback comes in the form of relativistic, collimated jets visible through synchrotron emission observed in radio wavelengths, which leave fingerprints on the ISM in the form of shock fronts and cavities perpendicular to the jet direction \citep{Wagner2011RELATIVISTICGALAXIES,Wagner2012DRIVINGINHOMOGENEITY}. This means that the orientation of the jet can have significant influence on the detectability and effectiveness of kinetic feedback \citep{Vayner2021AOutflows,Smirnova-Pinchukova2022TheTrends}. Radiative feedback, on the other hand, results from accretion onto the galactic nucleus leading to heightened UV radiation, which causes both the broad-line and narrow-line emission used to trace AGNs in optical wavelengths \citep{Heckman2014TheUniverse,Padovani2017ActiveName}. Radiation pressure and strong nuclear winds can power ionized and molecular outflows, resulting in shocks traced by X-ray emission \citep{Zubovas2013AGNGalaxies,Ishibashi2021AGN-drivenObservations,Veilleux2020CoolImplications}. 

Galaxy-scale outflows can result from both radiative and kinetic feedback processes, the extent of which has been well constrained by ionized outflow emission \citep{Cresci2015TheMUSE,Mingozzi2019TheMUSE,Revalski2018Quantifying573, Revalski2025QuantifyingSample}. Though molecular gas should dissociate at high velocities predominant in outflows, theoretically dense cold clumps of molecular gas can form and survive adjacent to hot gas \citep{Zubovas2014Galaxy-wideSpeeds}, explaining many of the molecular outflows identified near local AGNs \citep{Cicone2012The231,Cicone2014AstrophysicsObservations,Garcia-Burillo2014MolecularALMA}. Gas can be compressed along and at the peak of these massive outflows, leading to star formation activity known as positive AGN feedback \citep{Cresci2015TheMUSE,Maiolino2017StarOutflow,Gallagher2019WidespreadOutflows,Shin2019Positive5728}. But increased turbulence from outflows can also prevent the collapse of cold dense gas into stars, regulating star formation on longer timescales \citep{Choi2018TheGalaxies,Wylezalek2020IonizedNuclei,Gatto2024TheAGN}. Determining whether AGNs disperse cold gas (through removal or heating) or compress it and enhance star formation in a particular galaxy requires a careful census of the dense gas reservoir that fuels star formation activity. 

However, observational studies that consider the full dense gas disk of AGN hosts remain comparatively limited. Studies on the influence of AGNs on the densest parts of the ISM have so far been focused on the central few hundred parsecs composing the circumnuclear disk \citep{Usero2004Molecular1068,Aalto2012DetectionWind,Aalto2015ProbingHCN,Garcia-Burillo2021TheGATOS,Garcia-Burillo2024DecipheringGalaxies}, using molecular tracers such as HCN, HNC, and HCO$^+$ that should trace volume densities 10-100 times denser than CO \citep{Shirley2015TheTracers}. However, the conditions of the ISM can heavily influence the emission of these tracers, leading to misinterpretation of their abundance. HCN intensities, and HNC ones to a lesser extent, can be heightened by increased temperatures \citep{Hacar2020HCN-to-HNCISM,Harada2024AClouds}. X-ray irradiation in the vicinity of the AGN can heighten the electron abundance, leading to elevated HCN emission but a disruption of HCO$^+$ formation, affecting both emission and abundance \citep{Papadopoulos2007Galaxies,Zhang2014Dense6,Goldsmith2017ElectronRe-examined}. Indeed, many works point to heightened HCN emission in the centers of galaxies that host AGNs, bringing into question its validity as a dense gas tracer in galaxy centers \citep{Usero2004Molecular1068,Davies2012DenseNGC3227,Izumi2016DOHOLES,Imanishi2023DenseObservations}. More recent works demonstrate that HCO$^+$ emission may be a better dense gas tracer in spiral galaxy disks, exhibiting tighter trends with molecular gas surface density ($\Sigma_{\mathrm{mol}}$, \citealt{Stuber2025TheWhirlpool}), and less dependence on metallicity \citep{Patra2025VariationWay}. Within M51, HCO$^+$ emission is more tightly correlated with N$_2$H$^+$ emission as well, which should trace gas column densities greater than 10$^{22}$cm$^{-2}$, \citep{Pety2017TheGalaxies,Barnes2020LEGODisc,Tafalla2021CharacterizingCloud,Tafalla2023CharacterizingA}. Understanding how AGN feedback (whether in the form of outflows, shocks, or radiation) influences different dense gas tracers, and how far away from the AGN and any outflow these effects persist, is essential for both an accurate census of the dense ISM and for constraining the feedback processes that govern it.

M51, also known as the Whirlpool galaxy, is a perfect laboratory to probe the boundary between AGN feedback and normal star formation cycles near AGNs. M51 is a face-on, grand design spiral galaxy that has been observed at almost every wavelength, such that the multiphase components of AGN activity can be characterized. The galaxy itself is just the right mass ($7\times10^{10}$M$_{\odot}$, \citealt{Querejeta2015THEm}) for AGN feedback to be a predominant driver of star formation quenching \citep{Leslie2016QuenchingSequence, Lim2025InSimulation}, and the galaxy hosts a type II Seyfert nucleus \citep{Ho1997AGalaxies,Dumas2011THEM51}. The AGN itself, as identified by H$_2$O maser emission, exists at RA = 13:29:52.708, Dec = +47:11:42.810 \citep{Hagiwara2015HIGH-RESOLUTIONM51}, less than $\sim$0.1'' from a peak in the radio continuum \citep{Turner1994BrightGalaxies,Hagiwara2007Low-Luminosity4051}, with a black hole mass of $10^{6.96}\mathrm{M}_{\odot}$ \citep{Woo2002ACTIVELUMINOSITIES}. Radio observations also reveal a jet stemming from the AGN location inclined at 15 degrees with respect to the disk \citep{Cecil1988KINEMATICSM51,Dumas2011THEM51}, which itself has entrenched a molecular outflow containing both CO and HCN \citep{Scoville1998MolecularM51,Matsushita2007Special51,Matsushita2015RESOLVINGPUMPING,Querejeta2016AGNM51}. An ionized component with a wide (74 degree) opening angle was identified with the \textit{Hubble} Space Telescope \citep[HST;][]{Bradley2004PhysicalM51}, which coincides with a deficit in CO apart from an accumulation of clouds on the cone edge at $\sim$1'', possibly indicating localized positive feedback. Observations of the central 2 kpc using the IRAM Plateau de Bure Interferometer (PdBI) found enhanced and broadened HCN emission near the southern extranuclear cloud (XNC) noted by X-ray observations \citep{Querejeta2016AGNM51}. Though the jet is roughly only 4'' long (150\,pc) and 0.3'' (12\,pc) wide, evidence points to a more extended outflow emission in the form of this XNC and a larger radio lobe in the north as seen with the Very Large Array \citep[VLA;][]{Terashima2001Outflows}. Unlike multiple indicators of kinetic feedback, the AGN is not in a radiative efficient mode of feedback, based on {\it Chandra} and the Nuclear Spectroscopic Telescope Array (NuSTAR) observations that point to a Compton thick accretion with an Eddington ratio of $10^4$ \citep{Brightman2018ANuSTAR}.

In this paper, we combine the multiwavelength characterization of AGN feedback with new high-resolution results concerning the dense gas reservoir of M51 provided by Surveying the Whirlpool at Arcseconds with NOEMA (SWAN; \citealt{Stuber2025SurveyingSWAN}). SWAN maps a significant portion of the molecular gas disk of M51, rather than focusing on the nuclear region, allowing it to capture the boundary where AGN-governed feedback gives way to star formation. Though excellent work has been done to characterize HCN near the AGN, SWAN also provides alternative dense gas tracers such as HCO$^+$ and N$_2$H$^+$, which may not be as sensitive to nuclear activity. We investigate how AGN feedback influences these molecules to determine (A) under which conditions and environments accurate dense gas measurements can be made with specific tracers? And (B) to what extent AGN activity influences the dense ISM compared to other phases? We further consider the influence of shocks within M51, as the dynamical environment of this galaxy has been shown to create unique star formation depletion times \citep{Meidt2013GASPRESSURE,Colombo2014THEM51}. We achieve this by incorporating multiple optical diagnostics into our analysis, which can distinguish AGN ionization and star formation from low-ionization emission regions (LIERs) and shocks. By comparing optical diagnostics with X-ray observations characterizing the AGN activity, we create a complete picture of the feedback processes influence the dense molecular ISM.

In Sect. 2 we summarize the multiwavelength observations utilized in this work, and how different resolutions and fields of view are combined for our analysis. We also introduce the optical AGN diagnostics utilized in this work. We discuss our main results concerning the impact of AGN feedback on the validity of dense gas tracers, including correlations between molecular line ratios and multiwavelength tracers of AGN activity in Sect. 3. There we also consider the role shocks may play in modifying the ISM, both traced by ionized and molecular gas tracers. We discuss how our results compare to current theoretical and observational works investigating the dense ISM in Sect. 4, in particular previous results that use X-ray spectroscopy to constrain AGN feedback processes in M51. Lastly, we give some concluding remarks in Sect. 5.

\section{Methods}
\label{sec:methods}

M51 is one of the best-observed spiral galaxies in the local Universe, with large observational programs from the Very Large Array (VLA) \citep{Rots1990High-ResolutionM51,Dumas2011THEM51}, Institut de radioastronomie millimétrique (IRAM) / Northern Extended Millimeter Array (NOEMA) \citep{Schinnerer2013THEGALAXY,denBrok2022ACLAWS}, {\it Sophia} \citep{Pineda2018AFormation}, {\it Herschel} \citep{Schirm2017Probing5194}, HST (\citealt{Conroy2018ATimescales,Kessler2020Pa6946}), {\it Chandra} \citep{Terashima2001Outflows,Zhang2025FireM51}, and most recently a JWST treasury program. SWAN provides an essential additional view of the dense, molecular ISM at the scale of giant molecular clouds, probing the scales where low-luminosity AGN activity can influence the ISM and star formation process. We aim to select additional multiwavelength data products that complement the resolution and field of view achieved by SWAN. In the following section we summarize both new SWAN data products along with multiwavelength observations that provide further information on either the AGN or the ISM. Following this, we introduce our methodology to quantify the AGN impact on the ISM.

\subsection{Observations}

SWAN is an IRAM large program (LP003; PIs: F. Bigiel, E. Schinnerer) that maps multiple emission lines using the NOrthern Extended Millimetre Array (NOEMA) and the IRAM 30-meter telescope. These observations focus on the central 5\,kpc $\times$ 7\,kpc region of M51 to achieve a spatial resolution of $\sim$3'' ($\sim125$\,pc assuming a distance of 8.58$\pm$0.1 Mpc; \citealt{McQuinn2016THEM51}). Such a high resolution is required to resolve giant molecular clouds (see \citealt{Schinnerer2024MolecularGalaxies} and citations therein). This resolution is also necessary to resolve the radius of AGN impact, as other high-resolution AGN studies in the local Universe demonstrate that AGN impacts on the ISM can be limited to the central few hundred parsecs \citep{Garcia-Burillo2014MolecularALMA,Audibert2019Astronomy613,Esposito2022AGNLines}. Within this analysis, we focus on molecular tracers of dense gas probed in the 3\,mm band: HCN(1--0), HNC(1--0), HCO$^+$(1--0), and N$_2$H$^+$(1--0). For simplicity, we refer to the intensity of these line ratios as HCN, HNC, HCO$+$, and N$_2$H$^+$ from hereon. Though much fainter, SWAN also captures the slow shock tracer HNCO(4--3). A detailed description of the observational strategy and data reduction for SWAN is provided in \citet{Stuber2025SurveyingSWAN}.

SWAN is then combined with a variety of multiwavelength observations and data products using the \textit{PyStructure}\footnote{\url{https://pystructure.readthedocs.io/en/latest}} \citep{denBrok2022ACLAWS,Neumann2023TheALMA} code. \textit{PyStructure} convolves both 3D and 2D data to the same spectral (only relevant to the former) and spatial resolution, before measuring moment maps and their associated uncertainties for all 3D datacubes provided. We adopted the  \textit{PyStructure} configuration files utilized in previous SWAN publications (see \citealt{Stuber2023TheGalaxies,Stuber2025SurveyingSWAN}), using the 10\,km/s resolution datacubes publicly available on the IRAM data management system\footnote{\url{https://oms.iram.fr/?dms=showprograms}}. A 3D signal-to-noise ratio (S/N) mask was created using two priors: the HCN datacube, and $^{12}$CO(1--0) (CO from hereon) datacube from The PdBI Arcsecond Whirlpool Survey (PAWS; \citealt{Schinnerer2013THEGALAXY}), which laid the groundwork for SWAN. A detailed description of this masking, and the line distribution recovered by a dual-prior system, can be found in \cite{Stuber2025SurveyingSWAN}. Our work differs from previous SWAN studies once the data is convolved to a spatial resolution, for which we opt for a coarser resolution of 4.3'' rather than the native SWAN resolution of 3.05''. This is the coarsest resolution of the additional multiwavelength data products, all of which are combined into a single \textit{PyStructure} such that a consistent field of view and hexagonal pixel grid can be adopted. Below, we provide a brief summary of the additional data products and observations utilized in this work and integrated with \textit{PyStructure}.

\textit{HCN line widths:} In a companion paper, Usero et al. (in prep) built maps of several kinematic parameters by fitting the line cubes on a pixel-by-pixel basis with a single Gaussian profile. In particular, this provided them with linewidth maps that are more robust against noise at a low S/N position than the standard second-order line moments. In this work, we used the linewidth maps of the HCN line, which show the largest contrast between different regions at a S/N comparable to or better than CO(1--0). We also used the map of the reduced chi-squared statistic from the fit ($\chi_\nu^2$), which signposts the presence of non-Gaussian features. For further details and a more thorough analysis of the kinematics of the different lines observed by SWAN, see Usero et al. (in prep).

\textit{X-ray}: A full description of the {\it Chandra} X-ray observations and image processing is available in Rodriguez et al. (in prep). M51 was observed with {\it Chandra} ACIS-S for a total exposure time of $\sim$1.215~Ms. The datasets from the {\it Chandra} X-ray Observatory are contained in the {\it Chandra} Data Collection\footnote{\url{https://doi.org/10.25574/cdc.477}}, and were reprocessed and reduced with {\it Chandra} Interactive Analysis of Observations ({\sc ciao}) v4.17 \citep{Fruscione2006CIAO:System} and {\sc caldb} v4.12. The data were merged using \textit{merge\_obs} and point sources were detected using \textit{wavdetect} with a threshold of $10^{-6}$ and wavelet scales of $1$, $\sqrt{2}$, $2$, $2\sqrt{2}$, $4$, $4\sqrt{2}$, $8$, $8\sqrt{2}$, and $16$. Point sources were removed and filled using \textit{dmfilth}, and the merged observation was smoothed using \textit{aconvolve} with a $\sigma$ of 3 pixels along each axis to a resolution of 1.5''. However, point source removal is challenging in the center of M51, given that the X-ray emission is a combination of diffuse X-ray emission, unresolved point sources (and their extended emission), and emission from the AGN itself \citep{Kuntz2016AM51,Brightman2018ANuSTAR,Zhang2025FireM51}. Thus, we choose not to remove point sources in the central 50'' to avoid underestimating X-ray emission near the AGN. However, we have replicated this analysis with the central point sources removed, and find minimal differences in the qualitative trends as well as the computed correlation coefficients discussed in Sect. \ref{sec:correlations}. X-ray counts are then converted to an observed flux assuming a count rate - to - flux conversion factor of $4.281\times 10^{-17}$ erg/cm$^2$/s (accounting for a total exposure time of $1.215\times10^6$ s), and finally to luminosity by multiply by $4\pi D_{L}^2$ (where $D_{L}$ is the luminosity distance).

\textit{Star formation rate surface density}: Star formation rates were estimated using a combination of \textit{Spitzer} 24 $\mu$m maps \citep{Dumas2011THEM51} and HST H$\alpha$ maps \citep{Kessler2020Pa6946}, and the prescription provided in Eq. 6 of \citet{Leroy2013MOLECULARGALAXIES}. A more detailed description of the star formation rate surface density map, and its relation to different tracers in SWAN, can be found in \cite{Stuber2025TheWhirlpool}.

\textit{Radio continuum}: Radio observations at 6\,cm (4.9GHz) were collected covering the entirety of the galaxy disk using the VLA and short spacing corrections with the Effelsberg 100\,m telescope, a full description of which can be found in \cite{Dumas2011THEM51}. The radio continuum is strongest very close to the AGN \citep{Hagiwara2015HIGH-RESOLUTIONM51}, with an inner collimated jet extending $\sim$4'' to the south \citep{Crane1992THEM51} and a more extended radio loop extending $\sim$9'' to the north \citep{Dumas2011THEM51,Querejeta2016AGNM51}. The morphology traced by the radio continuum and X-ray emission are in excellent agreement, tracing a jet structure aligned with the ionized outflow characterized by the HST \citep{Querejeta2016AGNM51}.

\textit{VIRUS-P spectroscopy}: The central 4.1 $\times$ 4.1 kpc$^2$ of M51 were mapped using the Visible Integral field Replicable Unit Spectrograph Prototype (VIRUS-P) at the 2.7\,m \textit{Harlan J. Smith} Telescope at McDonald Observatory, with a spectral resolution of $\sim$5\,\AA\ full width at half maximum \citep{Blanc2009THE5194,Blanc2013THEMETHODS}. With a spatial resolution of 4.3'', VIRUS-P is the coarsest resolution data utilized in this study, and thus all other data products are convolved to this resolution ($\sim$180pc). Data reduction was performed as part of the The VIRUS-P Exploration of Nearby Galaxies (VENGA) survey using the VACCINE pipeline \citep{Adams2010THEGALAXIES}, a full description of which can be found in \cite{Blanc2009THE5194}. We focus our analysis on key emission line fluxes  H$\alpha$, H$\beta$, [OIII]$\lambda$5007 ([OIII] from hereon), and [NII]$\lambda$6584 ([NII] from hereon) along with velocity dispersions measured for the H$\alpha$ emission line. H$\alpha$ equivalent widths (EWs) are not provided as a data product from the VENGA survey and are thus were computed separately based on the final datacube using the procedures described in \cite{Westfall2019TheOverview} and also applied in \cite{Groves2023TheCatalogue}. To obtain them, we integrated the line flux within the rest-frame wavelength window 6557.6–6571.35 $\AA$. The continuum level was estimated by taking the median flux in two sidebands --- one on the blue side (6483–6513 $\AA$) and one on the red side (6623–6653 $\AA$) -- and then averaging the two median values. Finally, the H$\alpha$ EWs were estimated directly from the observed nebular spectrum, as the emission-line flux (the integrated flux within the line window) after subtracting the estimated continuum.

\subsection{BPT diagnostics and ELR function}
\label{sec:BPT}

Optical emission line ratios available from the VENGA data were used to quantify the influence of AGNs and shocks on the ISM, driven by either the nuclear outflow or other compressive forces in the galaxy. Baldwin-Phillips-Terlevich (BPT) diagrams \citep{Baldwin1981CLASSIFICATIONOBJECTS} have been effectively used to distinguish gas ionized by either star formation or AGN activity on both a global scale \citep{Baldwin1981CLASSIFICATIONOBJECTS,Kauffmann2003TheNuclei,Kewley2006TheNuclei,Stasinska2006Semi-empiricalHosts}, down to a kiloparsec scale \citep{Chen2019TheAGN,Lopez-Coba2019SystematicProperties,Pilyugin2020CircumnuclearDetectability,Kalinova2021StarGalaxies,Colombo2025TheEvolution} and now a sub-kiloparsec scale \citep{Husemann2022TheVariability,Smirnova-Pinchukova2022TheTrends,Watts2024MAUVE:Cluster}. Higher-resolution studies have revealed how the extent of AGN ionization can be greatly overestimated in kiloparsec-scale studies \citep{DAgostino2018StarburstAGNFraction}, with the highest resolution observations suggesting in some cases AGN ionization could be limited to tens of parsecs \citep{Nandi2023Evidence4395}.

\begin{figure}
    \includegraphics[width=0.99\columnwidth]{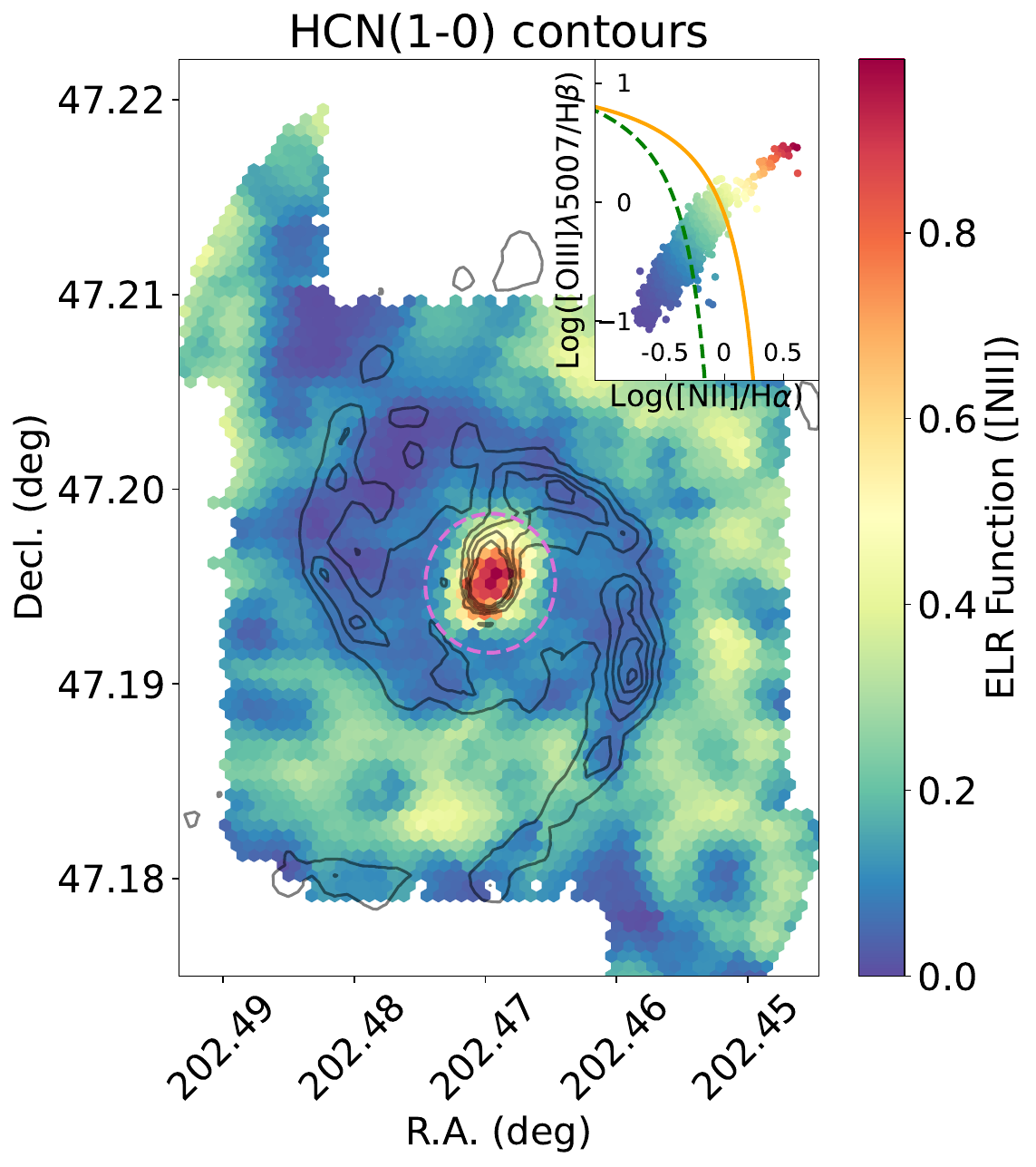}
    \centering
    \caption{Map of the ELR [NII] function over the SWAN field of view of M51, with HCN emission line contours shown in black for reference. The BPT diagram used to establish the ELR is shown in the top right, with values ranging from 0-0.25 for star-forming regions as defined by the \cite{Kauffmann2003TheNuclei} criteria (dashed green line) and values exceeding 0.5 for AGN regions (above the \cite{Kewley2001TheoreticalGalaxies} criteria, orange line). The central 0.5\,kpc is shown as a dashed pink line, to demonstrate how simple radial cuts to exclude AGN activity leave out the internal complexity of ionizing sources.}
    \label{fig:ELR_BPT}
\end{figure}

Traditionally, BPT diagrams have been used to identify a single ionizing source: HII regions (below the \cite{Kauffmann2003TheNuclei} line), AGNs (above the \cite{Kewley2001TheoreticalGalaxies} line), or a composite of the two (between the two lines). Figure \ref{fig:ELR_BPT} shows an example of a well established BPT diagram using these diagnostics, with \cite{Kauffmann2003TheNuclei} as a dashed green line and \cite{Kewley2001TheoreticalGalaxies} as a solid gold line. AGN ionization has often been further segregated into Seyfert emission (true AGN emission) or LIERs (\citealt{Kewley2006TheNuclei,Stasinska2008CanSurvey,CidFernandes2010AlternativeSDSS}). Older stellar populations in particular can lead to enhanced low-ionization line ratios utilized in these diagrams ([NII]/H$\alpha$ and [SII]/H$\alpha$) creating such LIER emission (e.g., \citealt{Belfiore2022AGalaxies}).

Rather than these strict demarcations, we adopted a metric that characterizes a smooth transition from HII region ionization to AGN ionization. We adopted a smooth transition that provides a unique value to all pixels based on the methods of \cite{DAgostino2019AData}. In this work an emission line ratio (ELR) function is established, which in combination with the H$\alpha$ velocity dispersion (H$\alpha$ $\sigma_{\mathrm{v}}$) and galactocentric radius (R$_{\textrm{gal}}$) creates a new 3D BPT diagram to isolate ionization from HII regions, AGNs, and shocks. The ELR function is smallest where HII regions dominate ionization, and highest where AGN ionization and/or shocks dominate emission. Values with high ELR and large H$\alpha$ $\sigma_{\mathrm{v}}$ at large radii are more likely the results of shocks than AGNs. The ELR function used in this work is defined as follows:

\begin{equation}
\label{eqn:ELR}
\begin{split}
    \textrm{ELR Function ([NII])} = \frac{\log(\mathrm{[NII]}/\mathrm{H}\alpha)-\textrm{min}_{\log(\mathrm{[NII]}/\mathrm{H}\alpha)}}{\textrm{max}_{\log(\mathrm{[NII]}/\mathrm{H}\alpha)} - \textrm{min}_{\log(\mathrm{[NII]}/\mathrm{H}\alpha)}} \\ \times \frac{\log(\mathrm{[OIII]}/\mathrm{H}\beta)-\textrm{min}_{\log(\mathrm{[OIII]}/\mathrm{H}\beta)}}{\textrm{max}_{\log(\mathrm{[OIII]}/\mathrm{H}\beta)} - \textrm{min}_{\log(\mathrm{[OIII]}/\mathrm{H}\beta)}}
\end{split}
,\end{equation}

\noindent where the minimum (min) and maximum (max) ratios are defined by the minimum/maximum value of all pixels in the map. This ELR function is based on the [NII] BPT diagram; however, the same methodology could be applied to any line ratio BPT replacing $\mathrm{[NII]}/\mathrm{H}\alpha$ with the $x$ axis ratios and $\mathrm{[OIII]}/\mathrm{H}\beta$ with the $y$ axis ratios. In Appendix \ref{app: alt BPT} we discuss how an alternative choice of BPT diagram could impact our results, but find that the main conclusions of this work remain unchanged.

Figure \ref{fig:ELR_BPT} shows a map of M51 where each pixel is assigned a value based on Eq. \ref{eqn:ELR}, along with the [NII] BPT diagram with the \citet[dashed green]{Kauffmann2003TheNuclei} and \citet[orange]{Kewley2001TheoreticalGalaxies} demarcation lines. All ELR values greater than 0.5 (yellow to red) lie above the \cite{Kewley2001TheoreticalGalaxies} line, and thus would be attributed to AGN ionization by traditional methods. Those with the lowest ELR values (blue) fall below the \cite{Kauffmann2003TheNuclei}, and indeed follow the structure of spiral arms and the molecular ring in agreement with older works \citep{Blanc2009THE5194}. Contours of the HCN emission traced by SWAN are shown for reference to highlight the agreement in structure. Unfortunately, the VENGA observations of M51 do not cover the full field of view as SWAN (particularly the edge of the spiral arms). The remainder of the analysis is limited to regions where ELR is measurable, and where SWAN lines reach a S/N greater than 3. All pixels within this field of view has a S/N greater than 3 for all optical emission lines used in the computation of the ELR. The SWAN S/N constraints result in the inter-arm regions being excluded from the analysis, with the center (including the nuclear bar), molecular ring, and spiral arms remaining. In some cases, we show for reference  SWAN emission lines both below this S/N cut and where the ELR is unmeasured, but we always distinguish these from the main analysis.

\section{Results}
\label{sec:results}
The ELR function is a powerful tool for quantifying the AGN contribution in individual pixels. We utilized this diagnostic and its complementary 3D BPT diagram to pursue three essential questions: whether the ELR can explain diverging dense gas relations identified by previous SWAN studies; whether the ELR predicts changes in the molecular ISM better than other tracers of AGN activity; and which signatures may be the result of mechanical feedback and shocks, as opposed to radiative feedback from AGN activity.

\subsection{Using ELR to identify regions where dense gas tracers diverge}
\label{sec:densegas_ELR} 
\begin{figure*}
    \includegraphics[width=0.95\textwidth]{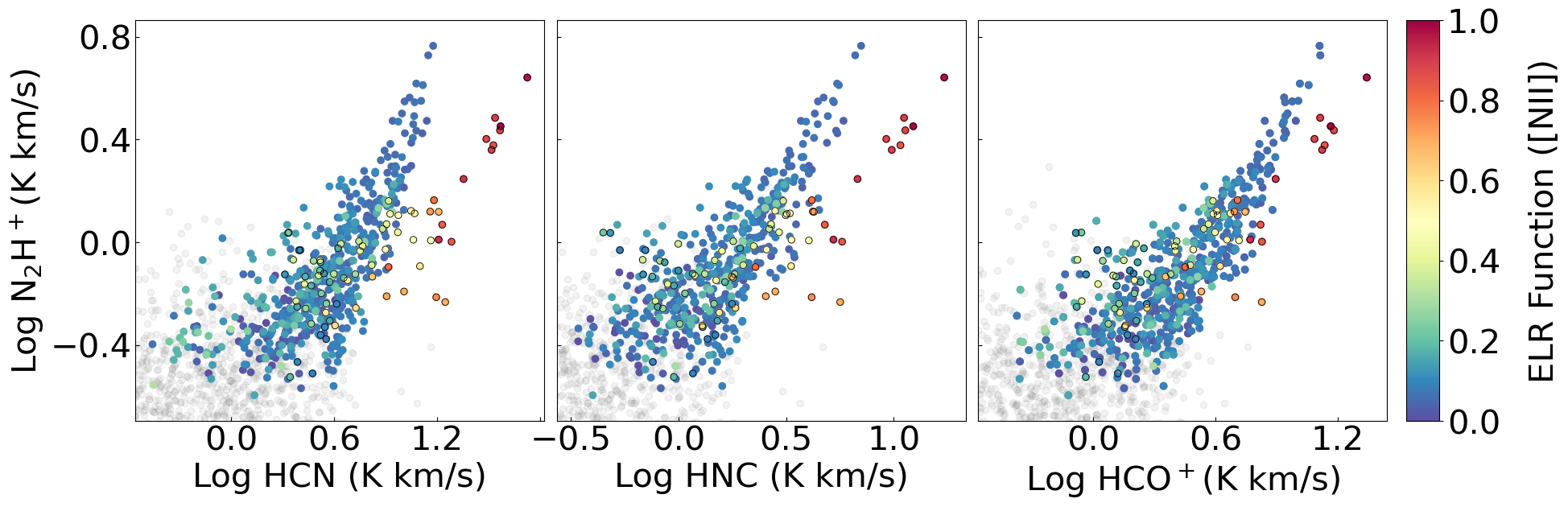}
    \centering
    \caption{Dense gas relations with N$_2$H$^+$, with alternative dense gas tracers on the $x$ axis, color-coded by the ELR function. Only points with a measurable ELR, which have a S/N greater than three for both lines, are included as colored points. Gray points are the remainder of pixels where the ELR is not measured, and the S/N of one or both emission lines is less than 3. The ELR accurately distinguishes ``problem regions'' where HCN and HNC no longer function as accurate dense gas tracers. HCO$^+$ is a better dense gas tracer, in agreement with N$_2$H$^+$ up to the most extreme ELR values. Yet even in this case, excluding ELR>0.75 would ensure the most accurate dense gas estimates, assuming N$_2$H$^+$ remains a viable dense gas tracer in the central regions. Pixels within the central 0.5\,kpc have a dark border, for spatial reference.}
    \label{fig:dense_gas_rel}
\end{figure*}

The initial SWAN publications noted that multiple dense gas tracers behaved oddly in the center of M51. In particular, HCN is dramatically excited compared to N$_2$H$^+$ in the central few hundred parsecs \citep{Stuber2023TheGalaxies,Stuber2025SurveyingSWAN}. Though molecules such as HCN, HNC, and HCO$^+$ have been used effectively to trace dense gas about to form stars across scales (e.g.,\ \citealt{Jimenez-Donaire2023AScales}), theoretical works warn that these molecules predominantly trace a transient over-density that will become diffuse before stars can form \citep{Priestley2023NEATHGas,Priestley2024NEATHCloud}. N$_2$H$^+$ is often considered a more accurate tracer of cold, dense gas that will be the sites of star formation in extragalactic targets. N$_2$H$^+$ is both destroyed in collisions with CO and dependent on the H$_3^+$ ion to form, which is also destroyed by CO interactions. Thus N$_2$H$^+$ only persists when CO is frozen to dust grains, tracing column densities above 10$^{22}$\,cm$^{-3}$ \citep{Kauffmann2017MolecularGalaxies,Tafalla2021CharacterizingCloud}. Ideally, comparatively easier to observe molecules such as HCN, HNC, and HCO$^+$ would have simple correlations with N$_2$H$^+$ so they could be used as proxy dense gas tracers. Figure \ref{fig:dense_gas_rel} shows how N$_2$H$^+$ emission correlates with these alternative dense gas tracers, color-coded by the ELR function. Within SWAN we see two separate correlations emerge, one for the majority of pixels with ELR$<0.5$, and one for pixels with ELR$>0.5$ where HCN, HNC, and HCO$^+$ are all brighter than expected for a given N$_2$H$^+$. This is not the result of lessened N$_2$H$^+$, in fact there is an excess of N$_2$H$^+$ emission with respect to CO emission near the AGN compared to other regions of the galaxy (see Fig. 5 of \citealt{Stuber2025TheWhirlpool}, also discussed in more detail in Sect. \ref{sec:discussion}). Nor is this the result of more significant line broadening in HCN, HNC, HCO$^+$ compared to N$_2$H$^+$; though there are variations in the ratio of line widths for all emission lines, these variations are mild compared to the change in integrated intensity ratios (Usero et al. in prep; see also Appendix \ref{app:LW})

Excess emission of alternative dense gas tracers could result from a plethora of physical phenomena. HCN, HNC, and HCO$^+$ are all more sensitive to the UV radiation field created by both high-mass star formation and the AGN than N$_2$H$^+$  \citep{Pety2017TheGalaxies}. Thus the ELR, which correlates with the radiation field from the AGN, would capture such regions. Additionally, within a strong IR radiation field as is present near an obscured AGN, HCN masing could be triggered by ``infrared pumping.'' Indeed, radiative transfer models applied to the center of M51 suggest a dense $\sim1$\,pc molecular disk around the AGN, leading to both HCN masing and nuclear obscuration \citep{Matsushita2015RESOLVINGPUMPING}. Enhanced cosmic rays from the AGN could also enhance HCN emission by heating the ISM; however, modeling of the kinetic gas temperature based on SWAN and Submillimeter Array (SMA) observations does not indicate a smooth increase in temperature toward the center of M51 \citep{denBrok2025COM51}. This does not mean that cosmic rays do not play a part in heightening HCN emission, but that multiple sources of cosmic rays (such as supernova remnants) likely play as important a role as AGN feedback.

If the cosmic ray ionization rate (CRIR) significantly influences molecular emission, a more dramatic change would be expected between N$_2$H$^+$ and HCO$^+$. An increased CRIR would lead to greater He$^+$ abundance, which reacts with and destroys CO to form HCO$^+$. N$_2$H$^+$ formation would also increase, from the destruction of CO as well as the production of H$_3^+$. Studies of clouds near the center of the Milky Way indicate N$_2$H$^+$ traces more extended molecular gas emission due to the increased CRIR in the galaxy center \citep{Santa-Maria2021SubmillimeterCores}, which could result in the tighter trend between N$_2$H$^+$ and HCO$^+$ (compared to HNC and HCN) seen in Fig. \ref{fig:dense_gas_rel}.

HCN, which exhibits the greatest offset from N$_2$H$^+$, could also be enhanced by X-ray radiation \citep{Maloney1996X-ray--irradiatedResults,Lepp1996X-ray-inducedClouds.,Tacconi1994The1068,Usero2004Molecular1068}. Within the central 0.5\,kpc, the ELR function is highly correlated with the X-ray luminosity (Kendall tau correlation coefficient $\tau_{\rm KT}=0.77$). The X-ray flux is mostly from mechanical feedback from the AGN, in the form of AGN winds and shocks with a small contribution from the circumnuclear disk (\citealt{Terashima2001Outflows}, further discussed in Sect. \ref{sec:shocks_xray}). But all of these photoionization sources, which lead to X-ray emission, will also result in large ELR values. Thus the ELR could capture regions where X-ray photons are most available to raise HCN emission. This would also explain why HCO$^+$ seems the least impacted, with only the central dozen pixels closest to the AGN (with the highest ELR) showing drastic difference in the relation.

Thus far we have only discussed how heightened emission of HCN, HNC, and HCO$^+$ might result in the dual trends seen in Fig. \ref{fig:dense_gas_rel}. However, the unique conditions of galaxy centers can also alter the gas conditions traced by N$_2$H$^+$, leading to this discrepancy. Both simulations \citep{Barnes2024CLOUDZONE,Petkova2023TheBrick} and observations \citep{Santa-Maria2021SubmillimeterCores} of the Milky Way central molecular zone indicate N$_2$H$^+$ traces a lower-density component of the gas reservoir than in the galaxy disk. Thus the ELR could also be a tool to identify regions where N$_2$H$^+$ traces a lower column density as a result of multiple properties unique to galaxy centers. In Sect. \ref{sec:correlations} we aim to disentangle the effect of AGN activity from general proximity to the galaxy center where molecular clouds are hotter, denser, and more turbulent \citep{Henshaw2020UbiquitousMedium}.

\begin{figure}
    \includegraphics[width=0.8\columnwidth]{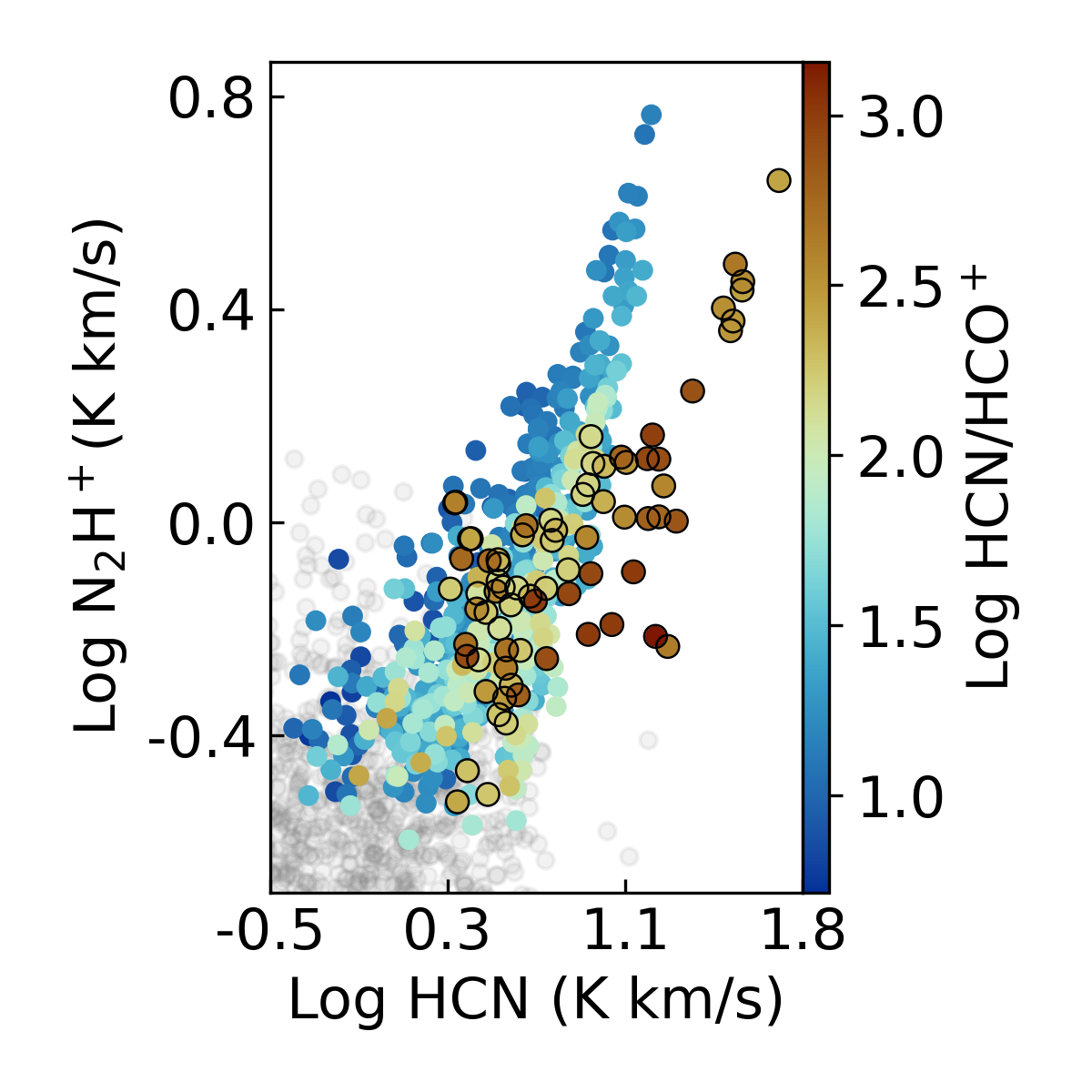}
    \centering
    \caption{Dense gas relation of N$_2$H$^+$ vs. HCN as shown in Fig. \ref{fig:dense_gas_rel}, but color-coded by HCN/HCO$^+$. This ratio is an alternative diagnostic for AGN contamination of molecular lines, with a ratio greater than 1.8 usually corresponding to obscured nuclear activity \citep{Kohno2005PrevalenceGalaxies}. Yet the main correlation between N$_2$H$^+$ (dominated by star-forming pixels) has many pixels with HCN/HCO$^+>$1.8. Indeed, the pixels with the largest HCN/HCO$^+$ ratio are not those with the greatest difference from the average trend, but those with a moderate difference but low N$_2$H$^+$ emission. Pixels within the central 0.5\,kpc have a dark border, to demonstrate how this ratio decreases far more gradually than the ELR function moving away from the galaxy center.}
    \label{fig:dense_gas_rel_HCNHCO}
\end{figure}

Regardless of the physical process dominating the ISM, the ELR captures subtle differences between  N$_2$H$^+$ and other dense gas tracers that would be missed by AGN diagnostics limited to SWAN lines. For example, HCN/HCO$^+$ has been traditionally used to distinguish X-ray-dominated regions (XDRs) and photo-dissociated regions (PDRs), often in comparison to HNC/HCO$^+$ or HNC/HCN \citep{Lepp1996X-ray-inducedClouds.,Kohno2005PrevalenceGalaxies,Meijerink2007DiagnosticsModels}. However we see a steadier increase in HCN/HCO$^+$ toward the center of the galaxy, and the largest ratio values are not coincident with the AGN (both based on position, and based on the ELR value.). Figure \ref{fig:dense_gas_rel_HCNHCO} demonstrates how selecting the largest values HCN/HCO$^+$ ratio would not select deviations in N$_2$H$^+$/HCN, unlike the HCN/HCO$^+$ cuts proposed by previous studies to distinguish obscured AGNs \citep{Kohno2005PrevalenceGalaxies}. Even if we consider HCN/HCO$^+$ in comparison to HNC/HCO$^+$ or HNC/HCN to diagnose XDR and PDR emission as in \cite{Baan2008DenseGalaxies}, all pixels in the SWAN/VENGA field of view fall into the PDR regime as was found in lower-resolution studies of M51 \citep{Eibensteiner2022A6946}. Though heightened HCN/HCO$^+$ ratios indicate the presence of an AGN, this ratio alone is sensitive to other mechanisms in the center of the galaxy that we discuss in more detail in Sect. \ref{sec:discussion}.

\subsection{ELR correlation with molecular tracers and other tracers of AGN activity}
\label{sec:correlations}

\begin{figure*}
    \includegraphics[width=0.99\textwidth]{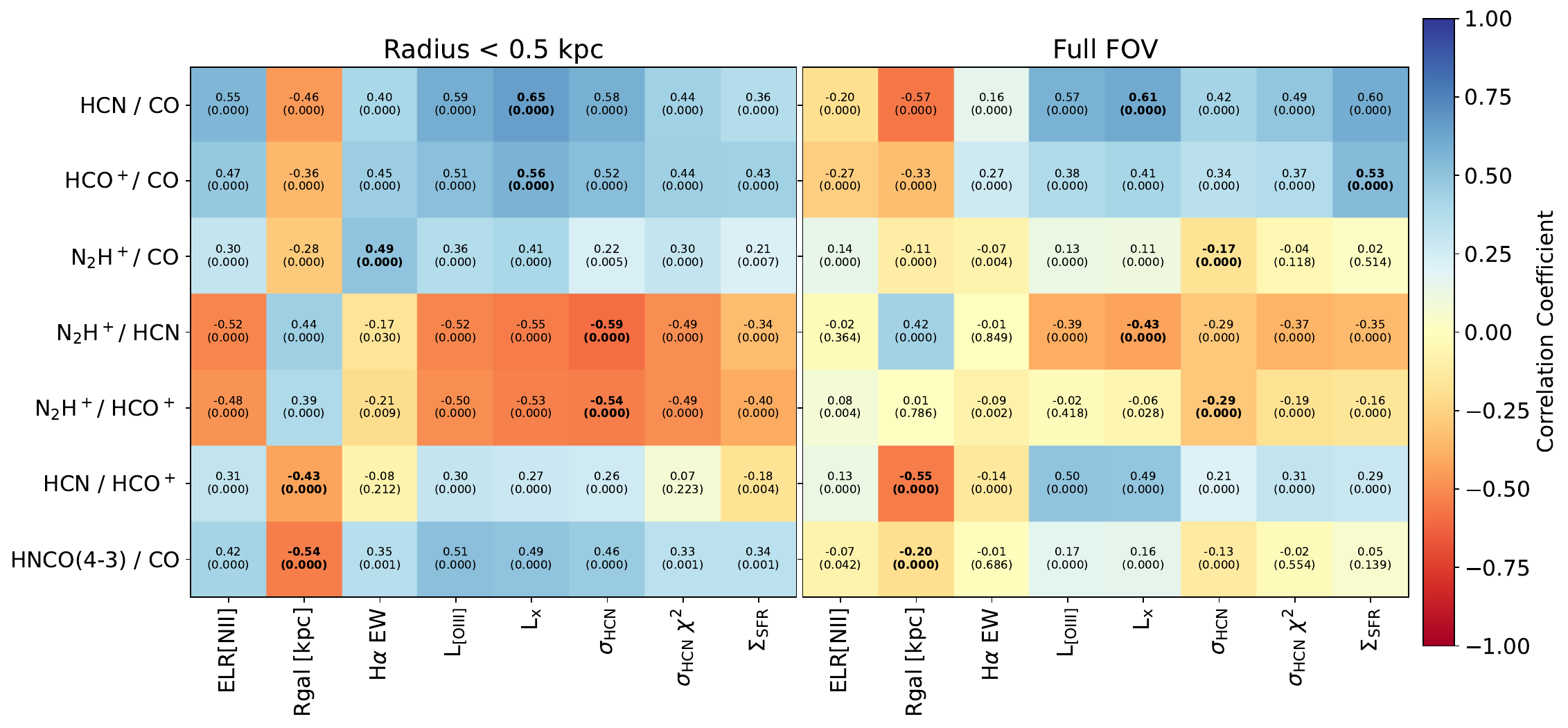}
    \centering
    \caption{Heat matrix for how well different tracers of AGN activity (x axis) correlate with line ratios from SWAN (y axis). Kendall tau correlation coefficients are shown in each box, and set the color scheme (with blue being highly correlated, red being highly anticorrelated, and yellow showing no correlation). $p$-values are shown in parentheses below. The strongest correlation for each SWAN line ratio is shown in bold. We consider both the central 0.5\,kpc (left), which should contain the observational indicators of the outflow and radio jet, as well as the full field of view (right). Variations in N$_2$H$^+$ emission compared to alternative gas tracers HCN and HCO$^+$ are always highly anticorrelated with tracers of AGN activity in the central 0.5\,kpc, with the strongest correlation being the Gaussian width of HCN. These correlations weaken when the full field of view is considered, but HCN/CO and HCO$^+$/CO correlations with AGN tracers remain strong. This points to AGN feedback (in its many forms) influencing the dense gas ratio throughout the SWAN+VENGA FOV, which in turn is highly correlated with $\Sigma_{\mathrm{SFR}}$.}
    \label{fig:correlations}
\end{figure*}

With the understanding that the ELR is connected to variations in the relations between different molecular emission lines, we test to see whether it is highly correlated with a range of SWAN line ratios. These ratios are chosen given their relation to the dense gas fraction (HCN/CO, HCO$^+$/CO, N$_2$H$^+$/CO), the validity of alternative dense gas tracers (N$_2$H$^+$/HCN, N$_2$H$^+$/HCO$^+$), photoionization sources in the ISM (HCN/HCO$^+$), or connections to shocks (HNCO(4--3)/CO). HNC is no longer considered to simplify discussion of these trends, as its relation to CO and N$_2$H$^+$ are qualitatively similar to those traced by HCN with respect to the ELR and other AGN tracers. Kendall tau correlation coefficients ($\tau_{\rm KT}$) along with their respective p-values between these ratios and the ELR are shown as the first column of the heat maps shown in Fig. \ref{fig:correlations}. We then directly compare these correlations to other properties that could be related to the strength of AGN feedback, including the distance from the galactic center in kiloparsecs ($R_{kpc}$), H$\alpha$ EW, $L_{\textrm{[OIII]}}$, $L_{X}$, and the width and $\chi^2$ to a Gaussian fit of the HCN line ($\sigma_{\mathrm{HCN}}$ and $\sigma_{\mathrm{HCN}}$ $\chi^2$, respectively). Though many of these are associated with AGN strength (H$\alpha$ EW, $L_{\textrm{[OIII]}}$, $L_{X}$), others are more strongly connected with outflows ($\sigma_{\mathrm{HCN}}$, $\sigma_{\mathrm{HCN}}$ $\chi^2$) and their subsequent shocks ($L_{X}$). We additionally provide $\tau_{\rm KT}$ for $\Sigma_{\mathrm{SFR}}$ to consider how the star formation radiation field may regulate particular molecular ratios, with the understanding that the combined 24 $\mu$m and H$\alpha$ estimator of star formation likely has contamination from AGN photoionization. A more detailed investigation into star formation rates in M51 and its influence on SWAN emission lines will be provided in Galić et al. (in prep).

As many of the effects of AGN feedback disappear beyond R$_{\textrm{gal}}$=0.5\,kpc (see also \citealt{Querejeta2016AGNM51}), we consider the inner 0.5\,kpc and the full field of view separately. For the common field of view shared between SWAN and VENGA, $\tau_{\rm KT}$ does not exceed an absolute value of 0.27 between ELR and any SWAN line ratios, indicating weak (anti)correlation between ELR and different line ratios. However this is somewhat expected, as the ELR is dominated by low values for the majority of the disk (see Fig. \ref{fig:ELR_BPT}). If we limit the correlation calculations to R$_{\textrm{gal}}$<0.5\,kpc, we find the ELR function has a strong, positive correlation ($\tau_{\rm KT}>0.3$) with HCN/CO, HCO$^+$/CO, N$_2$H$^+$/CO, HCN/HCO$^+$, and HNCO(4-3)/CO. Strong anticorrelations with N$_2$H$^+$/HCN and N$_2$H$^+$/HCO$^+$ also emerge in the center ($\tau_{\rm KT}<-0.3$).

However, ELR does not have the strongest correlation with millimeter line ratios compared to other properties that trace AGN feedback. For example, in the central 0.5\,kpc, HCN/HCO$^+$ is most strongly anticorrelated with  R$_{\textrm{gal}}$, and the $\tau_{\rm KT}$ value also increases in absolute value when the full field of view is considered. HCN/HCO$^+$ correlations with X-ray and [OIII] luminosity strengthen when the full field of view is considered, while the ELR correlation actually decreases. It is likely that HCN/HCO$^+$ is more tightly connected with the hot gas in the galaxy center, created by both AGN activity but also substantially by stellar clusters, and the anticorrelation with radius in the central 0.5\,kpc stems from a broader dependence on the ionized and/or hot gas radial distribution in the galaxy. 

In the central 0.5\,kpc, the outflow powered by the AGN becomes more important, with the strongest anticorrelations for N$_2$H$^+$/HCN and N$_2$H$^+$/HCO$^+$ being with $\sigma_{\mathrm{HCN}}$. Hypothetically, this correlation could result from N$_2$H$^+$ having a systemically narrower line width than HCN and HCO$^+$. In Appendix \ref{app:LW} we investigate how varying line width ratios could impact the integrated intensity ratios used here, to be sure they trace a true change in brightness as opposed to a change in line width.  We find only a few pixels where N$_2$H$^+$ line width is significantly smaller than HCN line width, ruling out such a scenario. Future SWAN publications will be devoted to a detailed comparison of the widths of different SWAN emission lines (Usero et al. in prep).

The deviations between different dense gas tracers correlating with $\sigma_{\mathrm{HCN}}$ indicate the outflow itself influences line emission, abundance, or possibly both. Both the high temperatures created on shock fronts as well as dust grain sputtering could lead to an overabundance of HCN especially \citep{Lefloch2021HCN/HNCASAI}. Additional evidence of the importance of shocks in the central 0.5\,kpc is the emerging strong correlations of HNCO(4--3)/CO with all tracers of AGN activity. HNCO(4--3) is a tracer of slow shocks which can result from AGN-triggered outflows \citep{Rodriguez-Fernandez2010AstronomyOutflow,Meier2015ALMA253,Kelly2017MolecularHNCO,Huang2023AstrophysicsALCHEMI}. Besides R$_{\textrm{gal}}$, HNCO(4--3)/CO has the strongest correlations with $L_{\mathrm{X}}$ and $L_{\mathrm{[OIII]}}$. In Sects. \ref{sec:shocks_optical} and \ref{sec:shocks_xray} we discuss how shocked gas contributes to the extreme emission in the center of M51, and how this helps us disentangle different feedback mechanisms powered by nuclear activity.

Despite the outflow, shocks, and/or AGN ionization governing N$_2$H$^+$ relations to other dense gas tracers, N$_2$H$^+$/CO is most strongly correlated with H$\alpha$ EW in the center. Young, massive stars and strong AGN ionization can both lead to large H$\alpha$ EW \citep{CidFernandes2010AlternativeSDSS}, so this could imply that N$_2$H$^+$/CO is also tracing star formation activity even toward the center of M51. However, the lack of correlation with $\Sigma_{\mathrm{SFR}}$ makes such a scenario unlikely. In the presence of heightened X-ray flux, we would expect lower N$_2$H$^+$/CO \citep{Meijerink2007DiagnosticsModels}, yet N$_2$H$^+$/CO has a strong positive correlation with $L_{\mathrm{X}}$ in the center (though not the full FOV). A more likely scenario would be a dense gas outflow, leading to heightened N$_2$H$^+$ abundance, along with HCN and HCO$^+$, compared to CO (which is low in the center of M51). A higher-resolution study of the central 0.5\,kpc would be necessary to fit Gaussian profiles to all dense gas lines (not just HCN) to better constrain what molecules are entrenched in such an outflow.

\subsection{Distinguishing AGNs and shocks with the 3D BPT}
\label{sec:shocks_optical}

Distinguishing AGN activity from shock activity with optical emission remains difficult. Shocked gas can produce large [OIII]/H$\beta$ and [NII]/H$\alpha$ ratios similar to that of AGN ionized gas \citep{Rich2011GALAXY-WIDEGALAXIES,Kewley2013THEOBSERVATIONS,Davies2014Starburst-AGNGalaxies,DAgostino2019Separating1068}. High-resolution long-slit spectroscopy performed with the HST along the radio jet of M51 does indicate both AGN and shock ionization, with the latter appearing to be the dominant source of [OIII] ionization south of the nucleus near the XNC (\citealt{Bradley2004PhysicalM51}). Shocks can be identified by high velocity dispersions of optical emission lines, ranging from $\sim$100-500 km/s as traced by ionized gas \citep{Rich2014CompositeShocks}.

The 3D BPT diagram proposed by \citet{DAgostino2019AData} accounts for shocks by including the velocity dispersion of the emission line, in this case the H$\alpha$ velocity dispersion H$\alpha$ $\sigma_{\mathrm{v}}$, as an additional axis. Together with R$_{\textrm{gal}}$ and the ELR function this makes for a three-dimensional diagram that can distinguish star formation, AGNs, and shocks. If, for example, ELR is high and H$\alpha$ $\sigma_{\mathrm{v}}$ is high, but values are far from the center, such regions are likely the sites of a shock front. If the ELR values are high and in the center, and H$\alpha$ $\sigma_{\mathrm{v}}$ is small, then AGN photoionization is the source of extreme line ratios. In the case where both are high in the center it becomes difficult to disentangle what is the result of shocks and what is the result of AGN photoionization alone. But combining this information with other tracers in SWAN can help break this degeneracy.

\begin{figure*}
    \includegraphics[width=0.99\textwidth]{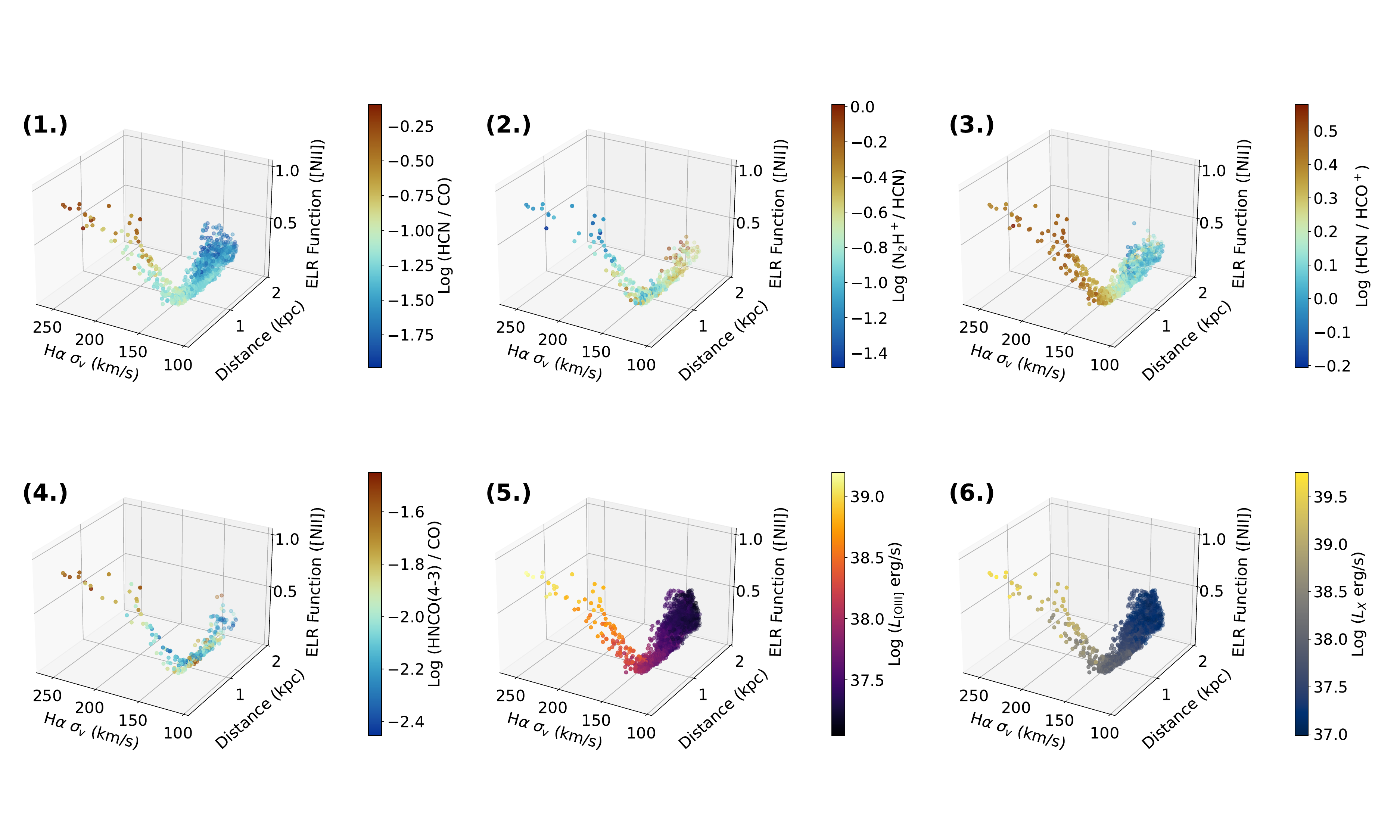}
    \centering
    \caption{3D BPT diagrams, with axes corresponding to the velocity dispersion of H$\alpha$ (H$\alpha$ $\sigma_{v}$ in km/s, the distance from the center of the galaxy in kiloparsecs, and the ELR function. Points are color-coded by relevant line ratios. \textit{Top:} (1.) HCN/CO ratios are largest where the ELR is largest, with a secondary dependency on H$\alpha$ $\sigma_{v}$, clear evidence of AGN feedback leading to excitation of HCN. (2.) N$_2$H$^+$/HCN ratios drop in the center, but only where ELR exceeds 0.5. These values also correspond to the highest velocity dispersion, meaning shocks cannot be ruled out as an ionizing source. But the lack of correlation between N$_2$H$^+$/HCN and velocity dispersion implies AGN ionization may play a bigger role. (3.) HCN/HCO$^+$, which should traditionally separate AGNs and star formation clearly fails compared to ELR, finding large values even as both ELR and velocity dispersion drop, and ionization from HII regions should dominate. \textit{Bottom:} (4.) HNCO(4--3)/CO, which traces slow shocks, is greatest where ELR and velocity dispersion are also at their highest, save a region near $\sim$1\,kpc where HNCO(4--3)/CO is large but ELR is low. (5.) [OIII] luminosity, which increases with both ELR and velocity dispersion, decreasing with distance. (6.) X-ray luminosity from {\it Chandra}, which exhibits similar trends as $L_{\textrm{[OIII]}}$. $L_{X}$ and $L_{\textrm{[OIII]}}$ increase more gradually with the ELR, similar to HCN/HCO$^+$, indicating they are sensitive to the overall interstellar radiation field and not AGN feedback alone.}
    \label{fig:3DBPT_shocks}
\end{figure*}

Figure \ref{fig:3DBPT_shocks} shows 3D BPT diagrams for the central 2\,kpc of the SWAN field of view. Panel 1 displays the 3D BPT with pixels color-coded by HCN/CO, which increases as ELR increases and radius decreases, with a secondary dependence on H$\alpha$ $\sigma_{\mathrm{v}}$. Comparisons of high-resolution HCN observations in the nucleus of M51 to high velocity outflows in Milky Way clouds point to HCN/CO ratios between 0.4 - 0.9 corresponding to shock-triggered values \citep{Matsushita2015RESOLVINGPUMPING,Umemoto1992TheGas,Wright1996AOrion-KL}, in agreement with the darkest brown points in Fig. \ref{fig:3DBPT_shocks}, panel 1. These extreme HCN/CO values have similarly high ELR values, but can have a range of H$\alpha$ $\sigma_{\mathrm{v}}$, implying some may be traced to shocks while others are the result of AGN photoionization alone.

Similarly, large N$_2$H$^+$/HCN values (panel 2) correlate with ELR but not necessarily H$\alpha$ $\sigma_{\mathrm{v}}$, implying both shocks and AGNs boost HCN emission. HCN/HCO$^+$, in contrast, remains high even as ELR values drop to star forming ones (ELR<0.5), with the highest values of the ratio actually corresponding to ELR values with comparatively lower H$\alpha$ $\sigma_{\mathrm{v}}$ (panel 3). It is clear here, as in the correlation coefficients, that HCN/HCO$^+$ is more tightly correlated with radius than AGN or shock ionization. The interstellar radiations field could thus play a more important role in driving HCN/HCO$^+$ than the source of photoionization. Indeed, [OIII] and X-ray luminosities also remain large as ELR and H$\alpha$ $\sigma_{\mathrm{v}}$ decrease, further confirming that heating of the ISM by star formation feedback predominantly drives variations in this ratio.

Within SWAN we have limited detections of HNCO(4--3), with continuous coverage only within the central 300\,pc and along the spiral arms. Slow shocks ($v<20$~km/s) can enhance low upper-state HNCO abundances including HNCO(4--3) and HNCO(5--4) \citep{Martin2008TracingRegion1,Martin2009PHOTODISSOCIATION253,Rodriguez-Fernandez2010AstronomyOutflow}. HNCO forms on the icy mantles of dust grains \citep{Fedoseev2015Low-temperatureAnalogues}, and can be thermally desorbed from the grain surface at high gas densities where gas and dust are decoupled ($n_{\mathrm{H}_{\mathrm{2}}}>10^{4}$cm$^{-3}$, \citealt{Kelly2017MolecularHNCO}). Though the HNCO(4--3) detections in SWAN are sparse, we can use them to compare to the shock diagnostics employed in the 3D BPT diagram, to best identify regions likely impacted by shocks (both those triggered by the AGN, and those independent from nuclear activity). Regions with the highest H$\alpha$ $\sigma_{\mathrm{v}}$ also have the highest HCNO(4--3)/CO(1--0), implying these central most regions are truly dominated by shocks resulting from a molecular outflow only marginally resolved by our investigation. If we consider the instrumental dispersion of the VENGA data ($\sim$123\,km/s), these regions would correspond to ionized shocks of $>$100 km/s, the minimum usual accepted value from optical shock studies \citep{Rich2014CompositeShocks}. It is possible both fast and slow shocks are present in these regions, but the resolution of this dataset blurs them together such that the millimeter and optical shock tracers agree. There is also a small region where HCNO(4--3)/CO(1--0) is heightened, but both the low ELR and H$\alpha$ $\sigma_{\mathrm{v}}$ in agreement with instrumental dispersion, meaning no fast shocks are present in the optical emission. We discuss the origin of these regions in the next section.

\section{Discussion}
\label{sec:discussion}

\begin{figure*}
    \includegraphics[width=0.99\textwidth]{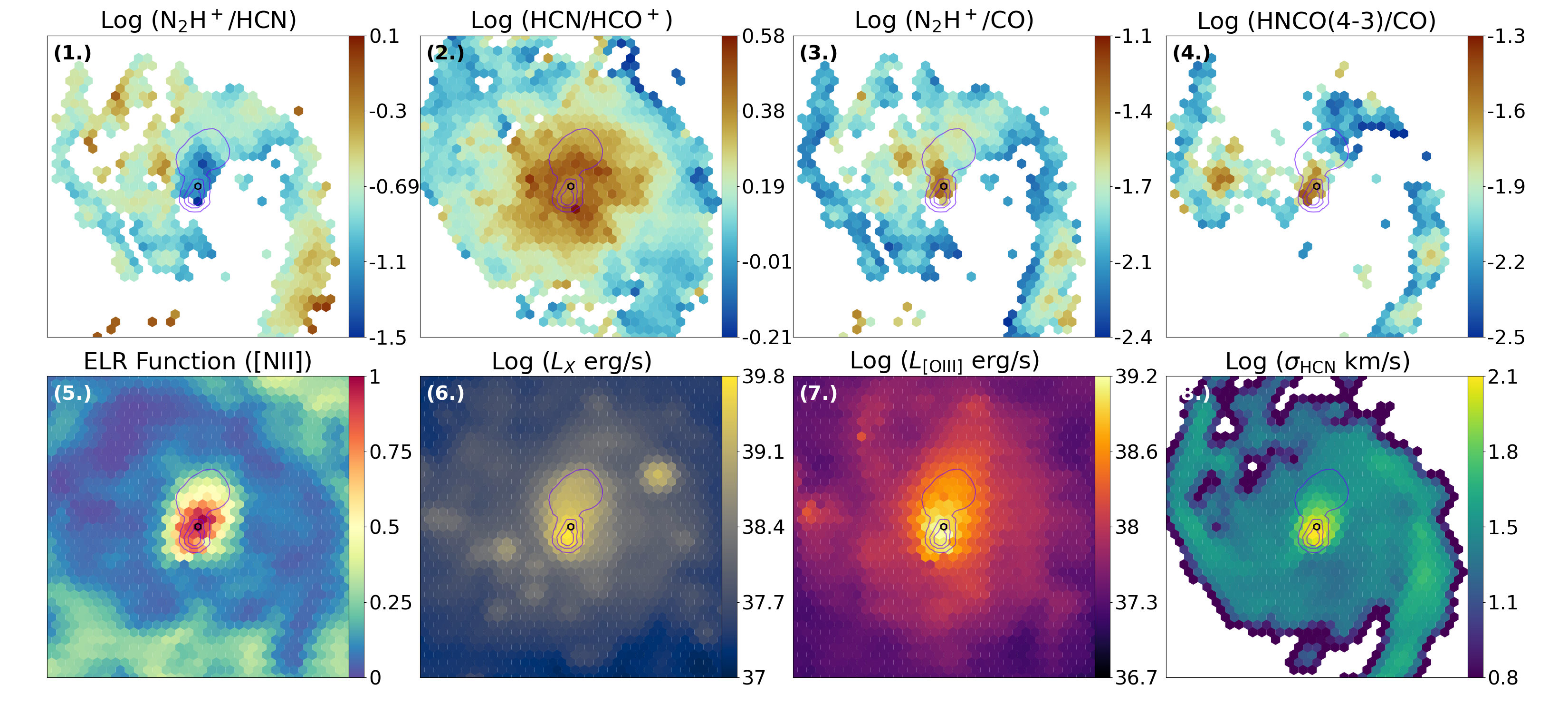}
    \centering
    \caption{Variety and extent of AGN impact as a function of the chosen indicator. Contours of the 6 cm radio continuum shown for comparison in purple. The pixel that hosts the AGN is circled in black, position provided by \cite{Hagiwara2015HIGH-RESOLUTIONM51}. \textit{Top:} (1.) N$_2$H$^+$/HCN ratio, demonstrating where the two dense gas tracers diverge (2.) HCN/HCO$^+$ ratio, traditionally used to distinguish PDRs from XDRs. Generally a ratio value greater than 1 (positive log-scale value) would indicate an XDR, though such values clearly exceed the AGN region here as defined by other indicators. (3.) N$_2$H$^+$/CO, which roughly traces the dense gas fraction, highlighting a potential dense gas outflow near the AGN (4.) HNCO/CO, a shock tracer with large values near the location of the AGN. Within central pixels VENGA resolution struggles to distinguish AGNs from shocks. \textit{Bottom:} (1.) ELR [NII] function, where orange-red points indicate AGN ionization dominates. (2.) X-ray luminosity from {\it Chandra}, note the heightened emission both along the radio lobe and particular south of the AGN. (3.) [OIII]5007 luminosity, with a clear ionized outflow extending north along the radio jet. (4.) The width of a Gaussian fit to the HCN emission line, note these values are highest near the AGN and more compact than all other AGN tracers. Future SWAN works will be devoted to investigating this region at higher resolution in more detail.}
    \label{fig:extent}
\end{figure*}

There is clear disparity in the tracers of AGN activity in the center of M51. Where different indicators agree, and where they diverge, can reveal new information about the physics at the center of this well-studied galaxy. Figure \ref{fig:extent} shows many of the AGN and shock tracers discussed so far, with the 6~cm radio continuum displayed as contours to indicate the location of the kiloparsec-scale radio jet. The pixel that contains the nucleus itself is circled in black, based on the right ascension and declination coordinates provided by \cite{Hagiwara2015HIGH-RESOLUTIONM51}. Though all AGN and shock indicators are highest in the center, closer to the southern radio peak and nucleus, their extent and morphology can vary dramatically. $L_X$ traces the radio continuum almost perfectly, whereas the ELR function has a more circular and symmetric distribution (with a slight lean toward the northern radio lobe). Indeed, the ELR is strongest toward the active nucleus, unlike excitation tracers  $L_{\mathrm{[OIII]}}$ and $L_{\mathrm{[X]}}$ which peak at the extra-nuclear cloud (XNC) in the south.

The ELR function provides the cleanest method of removing regions where N$_2$H$^+$/HCN diverges from the global power law. An X-ray luminosity threshold could be used, but no physically motivated luminosity cut emerges without under-sampling or over-sample the AGN ``contamination'' region. The Gaussian fit width of HCN could also be used, but risks selecting interesting dynamic regions beyond the nucleus, requiring additional radial cut for accuracy. An accurate but simple classification of AGN-contaminated regions is extremely important for the study of dense gas in galaxy centers. Averaged across spatial scales the N$_2$H$^+$-to-HCN ratio has a uniform scatter about the mean value of 0.15 \citep{Jimenez-Donaire2023AScales}, but the scatter in that value can be twice as large or twice as small. Understanding these variations is crucial to a growing field of resolved, extragalactic dense gas surveys, many of which are dominated by bright central HCN emission (e.g., \citealp{Neumann2023TheALMA}). Creating a way to systemically and uniformly correct for AGN contribution is essential, and the ELR offers a route that could be applied systematically across galaxy samples. The ELR serves as a robust indicator of regions where HCN is significantly brighter than N$_2$H$^+$, capturing all pixels with ratios less than 0.15 (0.82 in log scale, as shown in Fig. \ref{fig:extent}).

The ELR function, however, cannot explain regions where N$_2$H$^+$/HCN is uncharacteristically high, particularly left of the AGN region and in the southern arm. Although in the Milky Way N$_2$H$^+$/HCN varies between individual star-forming clumps, such variations would be averaged out by the SWAN beam size \citep{Feher2025DenseClumps}. More likely, larger scale galactic environment must be responsible for any noticeable deviations. \cite{Stuber2025TheWhirlpool} finds that R$_{\textrm{gal}}$, stellar mass surface density ($\Sigma_{\star}$), and dynamical equilibrium pressure ($P_{DE}$) all also play an important role shaping the relations between N$_2$H$^+$ and other dense gas tracers in M51. But most importantly, galactic gas dynamics such as those created by bars can create large-scale shocks that impact this ratio \citep{Usero2006Large-scale342,Meier2015ALMA253,Aladro2015LambdaGalaxies}. Gas dynamics could explain the heightened N$_2$H$^+$/HCN emission in the southern arm, which also corresponds to slow shocks traced by HNCO(4-3)/CO(1-0). Future SWAN studies will focus on the dynamical effects in these regions beyond the nucleus.

In agreement with previous SWAN publications \citep{Stuber2025TheWhirlpool}, this work points to HCO$^+$ is an essential alternative dense gas tracer, especially in galaxies with weak AGNs such as M51. Moderately high ELR values ($0.5<$ ELR $<0.7$) lead to dramatic shifts in HCN and HNC compared to N$_2$H$^+$. HCO$^+$ values with moderate ELRs are within the scatter of the N$_2$H$^+$- HCO$^+$ global power law. Unlike N$_2$H$^+$, HCO$^+$ is much quicker to detect in extragalactic sources, meaning it is an important compromise for a dense gas measurements in AGN hosts. The stability of HCO$^+$ emission near low-energy AGNs could be the result of its independence of the CRIR. An increased CRIR near the AGN boosts the ionization of H$_2$ and the subsequent increased free electron fraction would boost HCN emission \citep{Maloney1996X-ray--irradiatedResults,Papadopoulos2007Galaxies,Goldsmith2017ElectronRe-examined}. Though the over-abundance of H$_3^+$ in this environment could also lead to heightened HCO$^+$ abundance (e.g., \citealt{Harada2019ChemicalM83}), the increased free electron fraction would lead to the destruction of both H$_3^+$ and HCO$^+$ \citep{LePetit2016PhysicalH3+}, meaning HCO$^+$ emission would be constant with CRIR, but HCN/HCO$^+$ would increase \citep{Meijerink2006IrradiatedX-Rays,Meijerink2007DiagnosticsModels}. However, the ratio of HCN/HCO$^+$ remains high far beyond the areas flagged by the ELR (Fig. \ref{fig:dense_gas_rel_HCNHCO}).

As discussed in Sect. \ref{sec:correlations}, HCN/HCO$^+$ is more tightly correlated with $L_{X}$ and $L_{\mathrm{[OIII]}}$ when the full field of view is considered. This is reflected in Fig. \ref{fig:extent}, which shows a heightened $L_{X}$ and $L_{\mathrm{[OIII]}}$ values within the molecular ring but not beyond this or traced by the spiral arms. These regions appear to correspond to HCN/HCO$^+$>1, and they likely are not the result of AGN feedback or ionized shocks alone. The products of stellar clusters, in particular supernova remnants and massive X-ray binaries, could be heating gas beyond ELR>0.5, leading to heightened X-ray emission, as suggested by a recent study of the kiloparsec-scale variations in IR emission and {\it Chandra} X-ray observations \citep{Zhang2025FireM51}. 100 pc resolution or better is required for HCN/HCO$^+$ to accurately identify XDRs; otherwise, there is too much mixing with photoionization \citep{Butterworth2025AGalaxies}. Many previous observational studies have demonstrated that HCO$^+$/HCN, even in conjunction with other molecular line ratios, as a failed AGN diagnostic on kiloparsec and sub-kiloparsec scales \citep{Privon2020AGrowth,Li2021DenseTracers,Eibensteiner2022A6946}.

Though the ELR does an excellent job at selecting regions where dense gas tracers such as HCN are problematic, it fails to perfectly distinguish where shocks may impact the ISM. The compactness of HNCO(4--3)/CO compared to HCN/CO and the ELR would imply that shocks have a more limited impact extent, whereas energy injection from the AGN persists further out. To check if this is the result of the low detection rate of HNCO(4--3), we inspect maps of HNCO(4--3)/CO with a S/N cut of 1, and find a similarly compact region of emission (though a more continuous fall off as well). However, HNCO(4--3)/CO(1--0) is also large in a few peaks in the southern arm, and the region connecting the northern arm and the molecular ring. These regions also correspond to large N$_2$H$^+$/CO values, implying there could be some build up of dense gas resulting from slow shock fronts. These regions are not reflected in the ELR function, though they do have larger HCN velocity dispersions. The spectral resolution limit of VENGA makes it impossible to distinguish fast and slow shocks with the ELR, but it does appear to have a bias away from slow shocks as seen in the southern arm. There is some ambiguity with HNCO(4--3) as a slow shock tracer, as it will increase with respect to CO(1--0) as the dense gas mass increases, and could also increase due to high temperatures (even without the presence of shocks \citep{Kelly2017MolecularHNCO}. However, the large N$_2$H$^+$/CO would imply heightened temperatures are unlikely in the center. A higher-resolution study is needed to truly discern the competing forces of slow shocks, fast shocks, and possible heating from the AGN in the central 100 - 200\,pc where HNCO(4--3) is highest. However, we can use the additional X-ray data to confirm the role of shocks compared to radiative feedback resulting in AGN photoionization.

\subsection{Distinguishing AGNs and shocks with X-ray spectroscopy}
\label{sec:shocks_xray}

Preceding observations of M51 with {\it Chandra} have demonstrated that the XNC and northern loop are the result of mechanical feedback from the AGN, with the AGN-powered outflow shock-heating this gas and leading to heightened X-ray emission \citep{Terashima2001Outflows}. This initial work (TW2001 hereon) used a 15 ks observation in 2000 with ACIS-S to complete a detailed morphological mapping of the X-ray emitting nuclear region. As of now, there are $\sim$1.2 Ms of {\it Chandra} ACIS-S observation of M51, providing significantly higher-quality data for both imaging and spectroscopic investigations. Thus we elect to replicate their analysis to confirm what components of X-ray emission are due to shocks and thus mechanical AGN feedback, as opposed to radiative feedback often associated with optical AGN diagnostics. 

Spectroscopic analysis was done on the deepest observation of 189 ks (ObsID 13814). Following TW2001, we extracted the spectrum of the nucleus from a circular region of $1.5^{\prime\prime}$ radius. This yielded 1857 total net counts, compared to 242 total net counts in TW2001. The background spectrum was extracted from a region far from the nucleus. The background-subtracted spectrum was binned to have at least 10 counts per bin to facilitate $\chi^2$ minimization and was analyzed using XSPEC\footnote{https://heasarc.gsfc.nasa.gov/docs/software/xspec/}. 

M51 contains a Type-2 AGN in which the direct continuum as well as the broad emission line region is hidden from our line of sight. This is consistent with the X-ray spectral analysis of the nucleus by TW2001 who found that the direct X-ray continuum of the AGN was completely obscured. However, the spectrum shows strong soft X-ray emission, likely from the circumnuclear region. With this in mind, we modeled the 0.5--8 keV nuclear spectrum as a Compton-thick AGN ({\it pexmon} model in XSPEC with only the reflected continuum). The soft X-ray region was modeled as an optically thin thermal plasma in collisional ionization equilibrium (using an {\it apec} component with elemental abundance set to solar metallicity (Anders E. \& Grevesse N., 1989, Geochimica et Cosmochimica Acta, Volume 53). Thus the model was $tbabs *( tbabs * apec )+ pexmon)$ where the first {\it tbabs} parameter corresponds to the absorption by the ISM of our galaxy, the Milky Way, and the second corresponds to the absorption at the source. We found that a single temperature model was not adequate to fit the soft-band spectrum, so we added one more $apec$ which provided our best-fit model with $\chi^2=127$ for $93$ degrees of freedom. 

\begin{figure}
    \centering
    \includegraphics[width=\columnwidth]{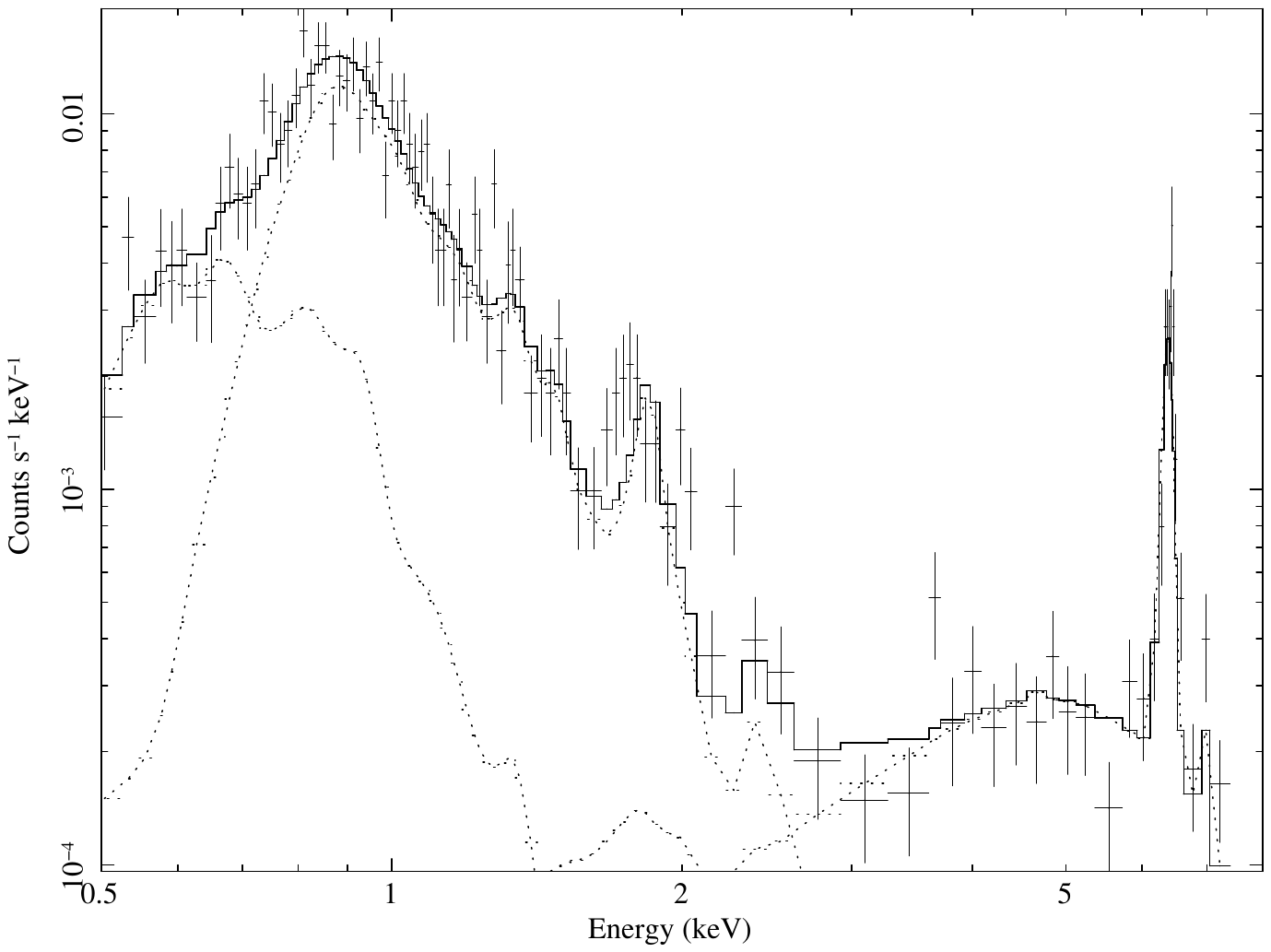}
    \caption{0.5--8 keV {\it Chandra} spectrum of the nucleus of M51 with the best-fit model. The dotted lines in the soft (0.5—2 keV) band show the two thermal components. In the hard (2--8 keV) band the strong Fe K$\alpha$ is seen which is the tell-tale sign of the reflection dominated continuum. The direct AGN continuum is completely obscured; the reflected continuum is shown with the dotted line. }
    \label{fig:xray_spectrum}
\end{figure}

In Fig. \ref{fig:xray_spectrum} we show the best-fit spectrum. We see strong thermal emission in the soft X-ray band. In the hard band, the direct power-law continuum is not visible at all, being completely absorbed. A strong Fe-$K\alpha$ emission line is clearly seen which is the tell-tale sign of the reflected spectrum. The best-fit parameters of the spectral fit are as follows. 
The soft X-ray spectrum is best fit by a two-temperature model with $ kT_{1}=0.15^{+0.02}_{-0.04}$ and $\rm kT_{2}=0.74^{+0.09}_{-0.13}$ keV. The photon index of the power-law continuum is not well constrained with $\Gamma=1.1^{+0.1}_{-1.1}$. 
The 0.5--2 keV flux is $2.96 \times 10^{-14} \rm erg s ^{-1} cm^{-2}$, and the luminosity is $2.6 \times 10^{38} \rm erg s ^{-1}$. In this 0.5--2 keV soft X-ray band, all the luminosity is from the thermal components. The hard X-ray luminosity in the 3--10 keV band is $9 \times 10^{38} \rm erg s ^{-1}$ and is entirely from the reflection component. Thus, at the nucleus soft X-ray are likely from the circumnuclear region while the hard X-rays are from the reflected continuum.  
% I"LL WRITE MORE NUMBERS HERE AND DOUBLE CHECK. 

Since hard X-ray emission is observed only from the nucleus (from \citealt{Hagiwara2015HIGH-RESOLUTIONM51}, marked in Fig. \ref{fig:extent}), we looked for morphological differences in molecular emission in the XNC, northern loop, and the nucleus as captured by X-ray emission and the radio continuum. The N$_2$H$^+$/HCN ratio has two distinct local minima aligned with the cloud and loop. Interestingly, this also aligns with the greatest $\sigma_{\mathrm{HCN}}$ at the southern cloud - all agreeing with a outflow scenario shocking the gas south of the nucleus. On the other hand, the ELR function has its peak at the AGN location, and not at the southern cloud or northern loop. Though the presence of hard X-rays is well correlated with the ELR, we do not necessarily expect hard X-ray flux to correlate with either high or low energy optical emission \citep{Berney2015BATLines,Agostino2023VLT-MUSELines,Pulatova2024WhatGalaxies}. Most theoretical work indicates X-ray emission captures more instantaneous accretion, whereas the narrow line region captured by the ELR traces an average accretion rate over much larger timescales \citep{Schawinski2015ActiveYr}. Rather the coincidence of the largest ELR values with hard X-ray emission from the nucleus highlight the potential of the ELR for identifying misclassified AGNs when X-ray observations are not available. Though to truly confirm its use in this regard, the ELR would need to be compared to similar X-ray spectral fitting for a broader range of both galaxy types and AGN strength.

These results are consistent with a ``two-stage'' AGN feedback scenario. In the first stage, the mechanical energy resulting from the AGN-powered jets heats the surrounding region to high enough temperatures that they emit in the soft X-ray band. In the second stage, these soft X-rays interact with and excite molecules such as HCN and HNC. However, there is a possible caveat. While the direct light from the AGN is obscured from our point of view, it may be visible to the cold gas that results in the molecular emission lines traced by SWAN. The fact that HCN and HNC emission are heightened more in the center where ELR is highest, and not along the southern XNC or the northern loop, lends merit to such a scenario. But we can also not rule out a possible heightened abundance of these molecules in the center due to a dense gas outflow. Beyond the unresolved jet-region ($<$150\,pc), a self-consistent picture of the nuclear region of M51 emerges where a combination of shock- and photon-energy results in heightened molecular emission.

\section{Conclusions}
\label{conclusions}   

SWAN has provided a new view into the chemistry and composition of the ISM in one of the best-studied spiral galaxies. We aim to supplement the extensive research analysis already performed on this dataset \citep{Stuber2023TheGalaxies,denBrok2025COM51,Stuber2025SurveyingSWAN,Galic2025SurveyingSWAN,Stuber2025TheWhirlpool} with a multiwavelength perspective, investigating gas-phases beyond the molecular to probe the boundary between AGN and star formation feedback processes.

Using integral field spectroscopy at a comparable resolution to SWAN, we defined an ELR function to differentiate regions dominated by AGN ionization (ELR>0.5) and star formation ionization (ELR<0.5), as shown in Fig. \ref{fig:ELR_BPT}. The ELR serves as the most robust metric of where alternative dense gas tracers such as HCN, HNC, and HCO$^+$ diverge from N$_2$H$^+$, leading to overestimates of the dense gas mass (assuming N$_2$H$^+$ is an accurate dense gas tracer). Though the overall scatter in the relations between N$_2$H$^+$ and alternative dense gas tracers is driven by physical variations in the ISM, more often tied to the stellar potential and molecular gas distribution \citep{Stuber2025TheWhirlpool}, an ELR<0.5 cut is essential to remove the obviously biased regions dominated by photoionization from AGN feedback. Even HCO$^+$, found to be the most consistent dense gas tracer with respect to $\Sigma_{\mathrm{mol}}$, R$_{\textrm{gal}}$, $\Sigma_{\star}$, and $P_{DE}$, has a select number of outliers beyond ELR>0.75, meaning the tracer is less impacted by AGN ionization in M51 but still should be treated with caution in the most extreme ionization regions. Importantly, ELR is only valid at capturing regions where alternative dense gas tracers are heightened with respect to N$_2$H$^+$; regions where N$_2$H$^+$ emission is enhanced are captured by other physical processes.

Increases in N$_2$H$^+$ line intensity appears to be tied to slow shocks, as traced by HNCO(4--3)/CO(1--0). Both fast shocks and slow shocks are likely present in the center of M51, as is demonstrated by the 3D BPT diagram (see Fig. \ref{fig:3DBPT_shocks}). A higher-resolution investigation of the center of M51 could begin to disentangle these processes, while also providing a more detailed kinematic analysis of the molecular outflow already noted in HCN and CO \citep{Matsushita2015RESOLVINGPUMPING,Querejeta2016AGNM51}. Such a study could provide novel insights into how N$_2$H$^+$ reacts to a complex kinematic and feedback environment.

The combination of millimeter, optical, and X-ray spectroscopy allows us to create a complete picture of the physical processes driving enhanced emission and abundance of different molecular tracers. Where ELR>0.5, AGN feedback processes prevail over star formation feedback, which governs the rest of the disk. Radiative feedback is likely present in the nucleus (marked in Fig. \ref{fig:extent}) at the highest ELR value, but the X-ray spectrum of the center of M51 indicates that the AGN continuum is otherwise completely obscured and that mechanical feedback dominates X-ray emission. Mechanical feedback is powered by the jet-ISM interaction, which leads to both soft x-rays and enhanced emission of dense gas tracers. Yet this jet can also have dense gas entrenched in a much smaller-scale outflow, as is implied by the heightened N$_2$H$^+$/CO and HNCO(4--3)/CO emission in the center. Given that only a dozen or so pixels pick up on this feature, the dense gas outflow will be investigated in future works with a higher spatial resolution toward the nucleus.

\begin{acknowledgements}
      This work made use of data from SWAN, the IRAM large program `Surveying the Whirlpool galaxy at Arcseconds with NOEMA'. This work was carried out as part of the PHANGS collaboration and is based on data obtained by PIs E. Schinnerer and F. Bigiel with the IRAM-30m telescope and NOEMA observatory under project ID M19AA. IRAM is supported by INSU/CNRS (France), MPG (Germany) and IGN (Spain). We'd like to thank the anonymous referee for their insightful feedback on this work. MDT thanks Kathryn Kreckel for advice on VIRUS-P observations of M51 and emission line measurements available through the VENGA collaboration. The construction of the Mitchell Spectrograph (formerly VIRUS-P) was possible thanks to the generous support of the Cynthia \& George Mitchell Foundation. We thank Phillip McQueen and Gary Hill for designing and constructing VIRUS-P, and for their advice on the use of the instrument. We also acknowledge David Doss and the staff at McDonald Observatory for their invaluable help during the observations. AU and MQ acknowledge support from the Spanish grant PID2022-138560NB-I00, funded by MCIN/AEI/10.13039/501100011033/FEDER, EU.
      DC, ZB, and FB gratefully acknowledge the Collaborative Research Center 1601 (SFB 1601 sub-project B3) funded by the Deutsche Forschungsgemeinschaft (DFG, German Research Foundation) – 500700252.  ES acknowledges funding from the European Research Council (ERC) under the European Union’s Horizon 2020 research and innovation programme (grant agreement No. 694343). LAL acknowledges support through the Heising-Simons Foundation grant 2022-3533. G.A.B. acknowledges the support from the ANID Basal project FB210003. JPe acknowledges support by the French Agence Nationale de la Recherche through the DAOISM grant ANR-21-CE31-0010 and by the Programme National ``Physique et Chimie du Milieu Interstellaire'' (PCMI) of CNRS/INSU with INC/INP, co-funded by CEA and CNES.
\end{acknowledgements}

\bibliographystyle{aa}
\bibliography{references} 

\begin{appendix}

\section{Alternative BPT diagnostics}
\label{app: alt BPT}
\begin{figure}
    \includegraphics[width=0.98\columnwidth]{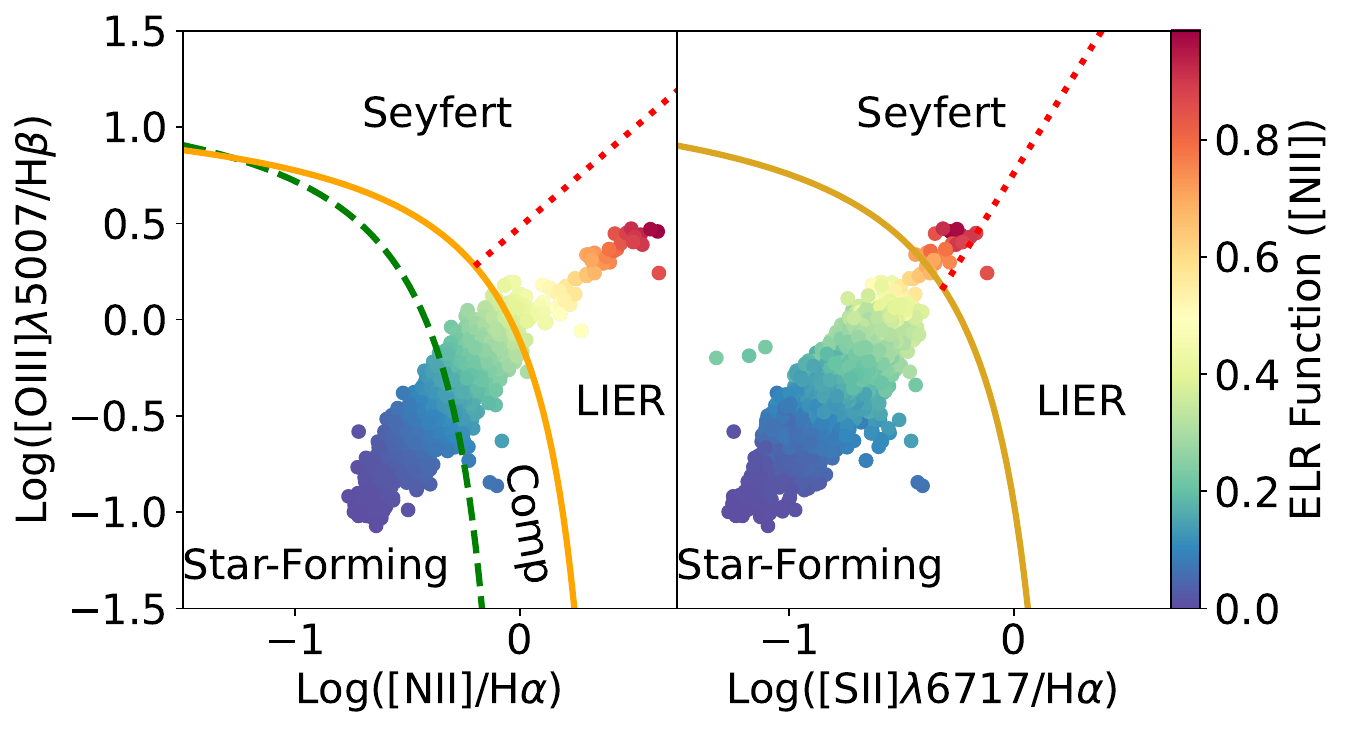}
    \centering
    \caption{Two alternative BPT diagrams to distinguish star formation from Seyfert and LIER emission, demonstrating how different BPT diagnostics may impact results presented here. All points are color-coded by the ELR function based on the [NII] BPT established in this work. \textit{Left:} BPT diagram using [NII]$\lambda$6584/H$\alpha$, similar to the one used in the main analysis but with additional separation of Seyfert and LIER emission. The \cite{Kauffmann2003TheNuclei} dashed green line selects for HII-region ionization, and the \cite{Kewley2001TheoreticalGalaxies} gold line selects AGN ionization. The pixels in between these two lines are considered ``composite.'' An additional distinction from \citet[dashed red line]{CidFernandes2010AlternativeSDSS} separates Seyfert ionization (top) from LIER emission (bottom). In the case of this BPT diagram, all ``AGN'' regions defined by the ELR would actually fall in the LIER regime, and thus are more likely to stem from shocks. \textit{Right:} BPT diagram using [SII]$\lambda$6717/H$\alpha$ instead of [NII]$\lambda$6584/H$\alpha$. Points below the \cite{Kewley2006MetallicityFlows} gold line stem from HII-region ionization, and Seyfert and LIER emission above this told line are again distinguished by the \cite{CidFernandes2010AlternativeSDSS} dashed red line. In the case of this BPT, in contrast to the left, all ELR>0.5 values but one fall into the Seyfert regime.}
    \label{fig:alt_BPT}
\end{figure}

Whether or not optical emission lines are dominated by AGN or shock ionization can also differ dramatically depending on the BPT diagnostic adopted. For simplicity this work has stuck to the same emission lines when measuring ELR, but we also consider how different BPT diagrams could impact our results. Using the same emission line ratios ([OIII], H$\beta$, [NII], H$\alpha$) we consider an alternative criteria to distinguish Seyfert emission from low-ionization emission regions (LIERs) proposed by \cite{CidFernandes2010AlternativeSDSS}, shown as a dashed red line on the [NII] BPT in Fig. \ref{fig:alt_BPT}. Here all points selected from the ELR in this (above the gold line) work would be considered LIER emission, which could stem from a low-energy AGN but are just as likely to come from shocks, or even photoionization from AGB stars. Another alternative is the [SII] BPT, which creates a similar separation between Seyfert and LIER emission but using the [SII]$\lambda$6717 emission line (right panel of Fig. \ref{fig:alt_BPT}, \citealt{Kewley2006TheNuclei}. In this case, the values with the most extreme HCN/CO, also those with ELR>0.5, are in the Seyfert region. If the [SII] BPT was adopted the AGN region established by the ELR would be far more compact, more confined to the nucleus and XNC than the full radio lobe. However, it would also agree better with HNCO(4--3)/CO(1--0) and the velocity dispersion of HCN, which should be regions where shocks dominate. 

There are clearly multiple physical processes at play even within a $\sim$180\,pc region of the galaxy, making a multiwavelength analysis essential to begin to disentangle these forces even if contradictions between tracers emerge. Including emission classified as a LIER by \cite{CidFernandes2010AlternativeSDSS} is important to account for the more extended shocks traced by X-ray emission that also influence emission lines in SWAN. The fact that the ELR function based on [NII] selects Seyfert regions both based on the \cite{Kewley2001TheoreticalGalaxies} [NII] and \cite{Kewley2006TheNuclei} [SII] BPT confirms our results are not dependent on the chosen optical emission lines.

\section{Line broadening and its impact on ratios of integrated intensities}
\label{app:LW}

Future SWAN studies will be devoted to a detailed analysis of broadening of various molecular lines, and their relation with AGN feedback in the center of M51 (Usero et al. in prep). However, given the dependence of this work on integrated intensity ratios, we aim to confirm that any trends with the ELR function (particularly those shown in Fig. \ref{fig:dense_gas_rel}) are not the result of the observed lines broadening to significantly different extents, leading to velocity-dependent variations in the line ratios. We thus test how the relationship between the ratio of integrated intensities for N$_2$H$^+$ and HCN compares to the ratio of peak temperatures (which should be insensitive to broadening effects) and line widths. We compute the line width for N$_2$H$^+$ and HCN using the approach adopted by \cite{Heyer2001TheGalaxy}, \cite{Leroy2016APROCESSES}, and \cite{Stuber2025TheWhirlpool}:

\begin{equation}
\label{eqn:lw}
    \sigma= \frac{I}{T_{peak}\sqrt{2\pi}}
\end{equation}

\begin{figure*}
    \includegraphics[width=0.98\textwidth]{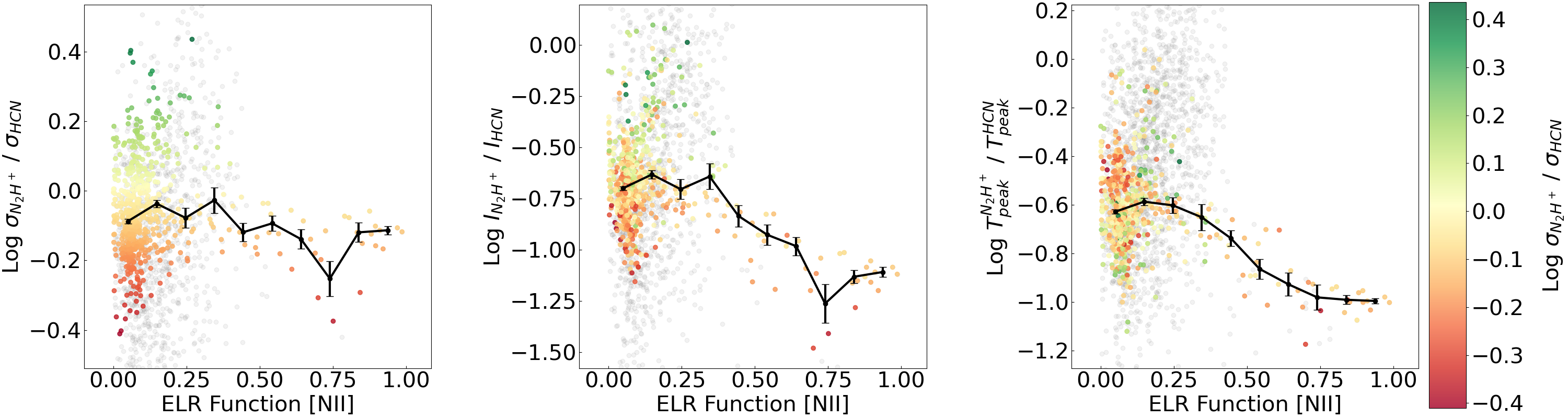}
    \centering
    \caption{Comparisons of how line width, line intensity, and peak temperature ratios vary with the ELR function. For every column, points are color-coded by the ratio of line widths for N$_2$H$^+$ and HCN, with green indicating N$_2$H$^+$ is the broader line and red indicating HCN is the broader line. Gray points indicate pixels where the S/N of one or both lines is less than three. The black points show the median y axis value in bins of ELR, with the error on the mean shown as a vertical bar. \textit{Left:} Ratio of line widths for N$_2$H$^+$ and HCN with respect to the ELR function. In general, this ratio has a flat trend with ELR, indicating that the central AGN regions do not pick out distinct regions where the broadening of these two lines differ. However, in the center of the galaxy where ELR is high, HCN is always slightly broader than N$_2$H$^+$. \textit{Middle:} Similar plot but with the ratio of integrated intensities on the y axis (the main method for determining line ratios in this work). As ELR increases the ratio of integrated intensities drops as HCN outshines N$_2$H$^+$. Importantly this trend is not entirely the result of HCN broadening compared to N$_2$H$^+$, as many red points (indicating a low ratio of line widths) are within the star-forming regime where ELR<0.5. \textit{Right:} Ratio of peak temperatures of these lines with respect to the ELR function. Peak temperature should be less impacted by line broadening, yet a clear trend of a decreasing ratio with increasing ELR persists.}
    \label{fig:lw_ELR}
\end{figure*}

\noindent Where $I$ is the integrated intensity of a line, and $T_{peak}$ is the peak temperature measured by the \textit{PyStructure} code. The first column of Fig. \ref{fig:lw_ELR} plots the ratio of line widths for N$_2$H$^+$ ($\sigma_{N_2H^+}$) and HCN ($\sigma_{HCN}$) against the ELR function, color-coded by this ratio. Though there are many regions in the galaxy where the line widths differ, with HCN usually being the wider of two lines, the ratio of line widths does not correlate with the ELR function ($\tau_{\rm KT}=0.07, \mathrm{p-value} = 0.0059$). There is a small dip in this correlation where the ratio of line widths drops dramatically, driven by less than a dozen pixels around ELR=0.75. Spatially this region is south of the AGN toward the XNC. This region will be the subject of study in Usero et al. (in prep), but contains so few pixels when data is convolved to VENGA resolution that its impact on the correlations and results presented in this work is relatively minor.

Important to note, however, is that regions where HCN is broader than N$_2$H$^+$ are not limited to this unique region at ELR=0.75. Many star-forming regions (ELR<0.5) have similar differences in line broadening, although a more careful analysis would be required to confirm these differences in such relatively faint regions. Despite this, the integrated intensity ratio is lowest at ELR=0.75, with significantly lower values than the star-forming regions with low line width ratios (red pixels in column 2 of Fig. \ref{fig:lw_ELR}). Similar conclusion are made if we replace the ratio of integrated intensities with ratios in $T_{peak}$, though the drop at ELR=0.75 is less dramatic. HCN has some additional broadening resulting from an unresolved hyperfine component only a few kilometers per second apart, yet if we replicate Fig. \ref{fig:lw_ELR} with HNC or HCO$^+$ instead of HCN, similar trends emerge.

In summary, though there are variations in the broadening of HCN and N$_2$H$^+$ emission lines, they do not appear to be the predominant driver of relations with the ELR function. Future works will be devoted to determining the physical conditions of regions where the lines differ, both related to the AGN (ELR>0.5) and in the galaxy disk (ELR<0.5).

\end{appendix}

\end{document}